\documentclass[12pt,preprint]{aastex}

\usepackage{graphicx}

\def\chandra{{\em Chandra}}
\def\xmm{{\em XMM-Newton}}

\def\xs{{N11}}
\def\loglxlbol{$\log(L_{\rm X}/L_{\rm BOL})$}
\def\lxlbol{$L_{\rm X}/L_{\rm BOL}$}

\shorttitle{Chandra observation of N11}
\shortauthors{Naz\'e et al.}

\begin{document}

\title{A Deep {\em Chandra} Observation of the Giant \ion{H}{2} Region N11 I. X-ray Sources in the Field}

\author{Ya\"el Naz\'e\altaffilmark{1}}
\affil{GAPHE, Department AGO, Universit\'e de Li\`ege, All\'ee du 6 Ao\^ut 17 Bat.\ 
B5C, B4000-Li\`ege, Belgium}
\email{naze@astro.ulg.ac.be}

\author{Q. Daniel Wang}
\affil{Department of Astronomy, B619E-LGRT, University of Massachusetts, Amherst, MA 01003, USA}

\author{You-Hua Chu, Robert Gruendl}
\affil{Department of Astronomy, University of Illinois, 1002 West Green Street, 
Urbana, IL 61801, USA}

\and

\author{Lida Oskinova}
\affil{Institute for Physics and Astronomy, University of Potsdam, D14476 Potsdam, Germany}

\altaffiltext{1}{Research Associate FRS-FNRS}

\begin{abstract} 
A very sensitive X-ray investigation of the giant \ion{H}{2} region N11 in the LMC was performed using the {\em Chandra X-ray Observatory}. The 300\,ks observation reveals X-ray sources with luminosities down to 10$^{32}$\,erg\,s$^{-1}$, increasing by more than a factor of 5 the number of known point sources in the field. Amongst these detections are 13 massive stars (3 compact groups of massive stars, 9 O-stars and one early B-star) with \loglxlbol $\sim-6.5$ to $-7$, which may suggest that they are highly magnetic or colliding wind systems. On the other hand, the stacked signal for regions corresponding to undetected O-stars yields \loglxlbol $\sim-7.3$, i.e., an emission level comparable to similar Galactic stars despite the lower metallicity. Other point sources coincide with 11 foreground stars, 6 late-B/A stars in N11, and many background objects. This observation also uncovers the extent and detailed spatial properties of the soft, diffuse emission regions but the presence of some hotter plasma in their spectra suggests contamination by the unresolved stellar population.
\end{abstract}


\section{Introduction}

With the decade-long work of sensitive X-ray observatories \xmm\ and \chandra, a refined picture of stellar X-ray emission in our Galaxy is now available \citep[for a review, see e.g.,][]{gue09}. However, many X-ray production processes in stars depend on metallicity, and this dependence has not yet been tested thoroughly. In this context, the Magellanic Clouds provide an opportunity to observe the metallicity effects. For example, the tentative discovery of X-ray emission from low-mass pre-main sequence (PMS) objects in the Small Magellanic Cloud (SMC) was reported recently and the emission level appears comparable to that in the Galaxy, constraining emission models \citep{osk13}. On the high-mass end, a 26\,ks \chandra\ exposure of the emblematic \ion{H}{2} region 30 Doradus was reported by \citet{tow06a,tow06b}, leading to the detection of 180 X-ray sources - a hundred being found in the massive central cluster R136, with some of these sources displaying $\sim$2 net counts and thus having a non-negligible probability to be spurious. From their spectral analyses, \citet{tow06b} derived for the massive stars of 30 Dor absorption columns of 1--10$\times 10^{21}$\,H-atom\,cm$^{-2}$, temperatures of 0.5--4\,keV, and absorption-corrected X-ray luminosities from $2\times 10^{35}$\,erg\,s$^{-1}$ down to $10^{33}$\,erg\,s$^{-1}$ (the sensitivity limit). \citet{tow06b} suggested that many, but not all, of the detected massive stars were colliding-wind binaries, which could be ascertained through further monitoring. Furthermore, these authors found no clear \lxlbol\ ratio for their sample, contrary to that for the Galaxy \citep[$\sim 10^{-7}$, see e.g.][and references therein]{naz09}. This conclusion may be reconsidered, however,  because (1) a significant contamination by the X-ray bright colliding wind binaries (esp.\ WR+O) may hide trends intrinsic to individual massive stars (but correcting this problem needs both optical and X-ray extensive monitoring, which is not available) and (2) coherent \lxlbol\ ratios are found only when X-ray luminosities are well measured statistically and are corrected by the {\it interstellar} absorption, not by the {\it total} absorption. It should also be underlined that 30 Dor is an extreme environment, more akin to starbursts than a good representative of the Large Magellanic Cloud (LMC) population.

In the LMC, N11 is the second largest {\ion{H}{2}}
region, just after the giant \ion{H}{2} region 30 Dor. 
The less extreme properties of N11 make it much more representative of LMC {\ion{H}{2}} regions and clusters of massive stars. Besides, the lower concentration of stars implies less source confusion, hence should lead to more reliable results.

Star formation has been
very active in N11, with no less than four recognized OB associations: LH9, 
LH10, LH13, and LH14 \citep{luc70}. The stellar feedback has restructured the 
surrounding interstellar material. Notably, the winds and supernova 
explosions of the massive stars in LH9 have gradually carved 
a cavity, giving rise to an expanding superbubble some 120 pc in size 
\citep{ros96}. This expansion has probably triggered the formation of 
LH10 at the periphery of the superbubble \citep{wal92}. In turn, the massive 
stars of LH10 are now beginning to blow bubbles \citep{naz01}, further 
triggering new star formation in their surroundings \citep{bar03}.

N11 is clearly one of the best sites to study the interplay between stars 
and the interstellar medium. This interaction, often violent, produces X-ray 
emission. Using a 30 ks {\it ROSAT} Position Sensitive Proportional
Counter  observation, \citet{mac98} reported the first detection of 
X-rays in N11. This {\em ROSAT} observation revealed the presence of 
extended areas of diffuse emission, with the brightest sources associated 
with the N11L supernova remnant (SNR) and within the superbubble around LH9. 
Further investigation was performed with a 30 ks {\em XMM-Newton}
observation, which provided the first detection of point sources in the field 
\citep{naz04}. While stars in LH9 remained unresolved, this {\em XMM} 
observation unveiled in LH10 a mixture of diffuse emission and point sources 
associated with some of the most massive stars of the cluster.

A detailed X-ray analysis of N11 requires a combination of both high
sensitivity and high spatial resolution, which became possible with our 
new, deep X-ray investigation of N11, using the {\em Chandra X-ray 
Observatory}. This observation will lead to several analyses, and this 
first paper discusses the point source population. The aim is to uncover 
the nature of the point sources, and to find whether the properties of 
the stars in N11 differ from similar objects in the Galaxy.
This paper is organized as follows: Section 2 presents the data and their 
reduction, Section 3 introduces the catalog of X-ray sources and 
their global properties, Section 4 discusses extended and point sources 
associated with massive stars.  Finally, Section 5 summarizes our
results.

\section{Observations and Data Reduction}\label{sec:data_red}

The {\em Chandra} ACIS-I observations of N11 were made in six separate 
segments within two months in 2007.
As summarized in Table\,\ref{obs}, the exposure time of each segment
was 42--49\,ks and the roll angle ranged from 130$^\circ$ to 188$^\circ$.
The pipeline products of the observations were reduced and analyzed 
using our own IDL-based analysis tools (e.g., \citealt{wan04a}) as well
as the official software for the \chandra\
Interactive Analysis of Observations (CIAO; version beta1-4.0) 
together with the calibration database (caldb 3.4.0) and other publicly
available routines (e.g., XSPEC v 12.7.0). 
It should be noted that our detection procedure was shown to give results 
comparable to those of ACISextract \citep{joh13}, and this was checked 
on this dataset by a quick run of that tool.

We used the light-curve cleaning routine ``lc$\_$clean''
to remove time intervals of significant background flares when count rates
deviated more than 3$\sigma$ or a factor of $\gtrsim 1.2$ from the mean 
rate of individual observations. This cleaning, together with a correction 
for the dead time of the six observations, resulted in a total of 280\,ks 
useful exposure for subsequent analysis.

A combination of source detection algorithms (wavelet, sliding-box, and maximum likelihood centroid fitting) were applied to unsmoothed data in three bands: soft (S) 0.5--2\,keV, hard (H) 2--8\,keV, and total (T) 0.5--8\,keV \citep{wan04a}. Briefly, the wavelet detection was first used to find the initial source candidates with a high threshold of local false detection probability $P < 10^{-5}$.  Then, background maps were constructed by removing the wavelet-detected sources and by conducting a median averaging or smoothing of the three input images on scales much greater than the point spread function (PSF) to achieve an intensity uncertainty $\sim 10\%$. These background maps, insensitive to the exact details of the construction procedure, were used to search again for sources, using this time a sliding-box algorithm (the so-called map detection mode). Finally, a maximum likelihood centroiding algorithm was used, still using the count background map, to derive the best centroid positions for the sources \citep{wan04a}. 

This process was applied independently to each energy band. Our final source list contains sources with local false detection probability $P<10^{-6}$ at least in one band (Poisson statistics was used in calculating the significance of a source detection above the local count background). The sensitivity of the source detection depends on the size of the PSF as well as the local background level and effective exposure, which all vary with position, especially with the off-axis angle of the detected sources. The source detection, though optimized for point-like sources, includes a few strong peaks of the diffuse X-ray emission, chiefly associated with the SNR N11L, about $\sim 7^\prime$ west of the field center (for more on this object, see Sun et al. in preparation). Some of these sources associated with peaks of the diffuse emission are detected in S-band only. 

\begin{figure}
\includegraphics[height=7in,angle=180, clip=]{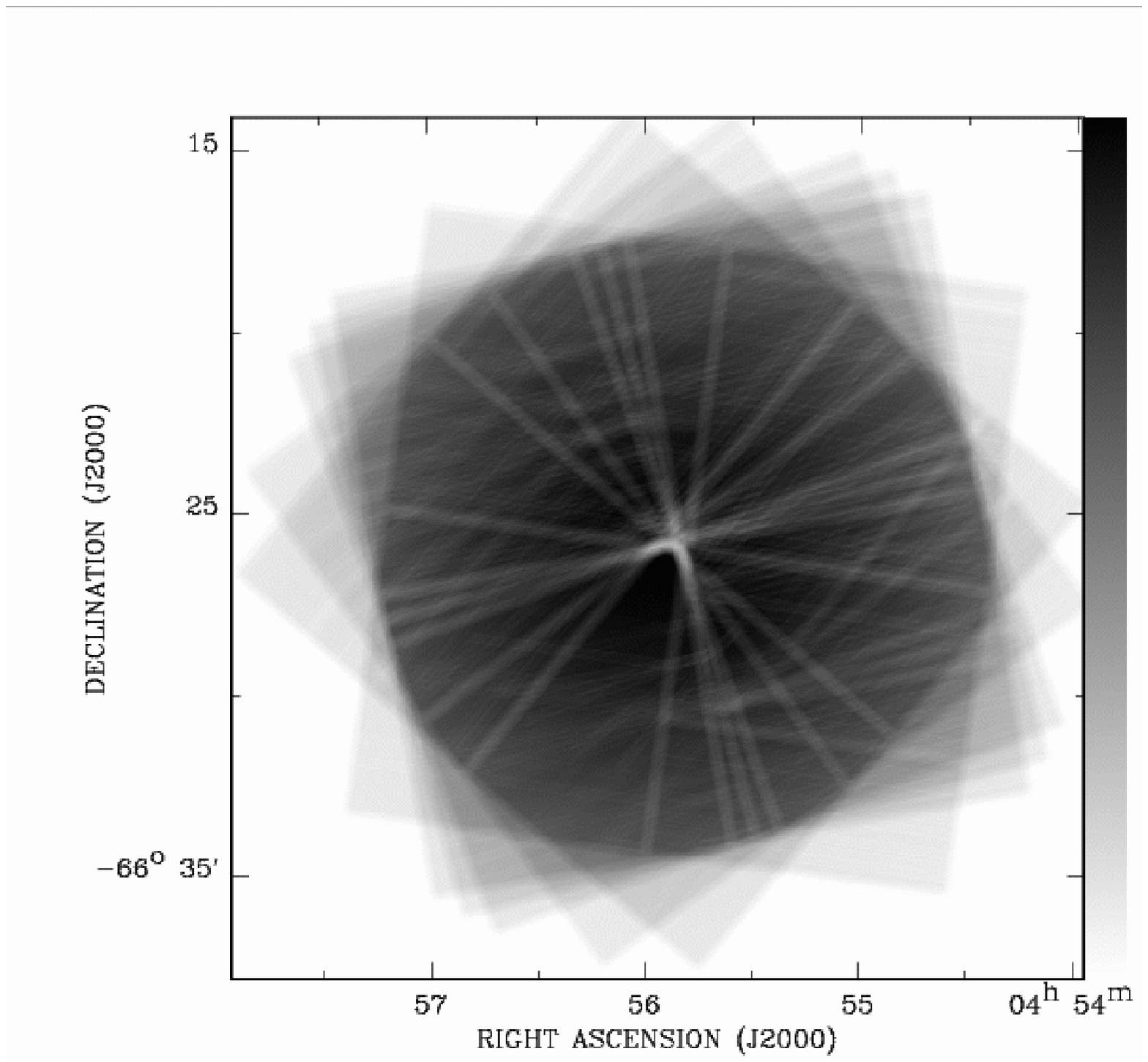}
\caption{{\bf ONLINE MATERIAL} Merged ACIS-I effective exposure map in the 0.5--1\,keV band. The exposure is linearly scaled in gray.} \label{fig:exp_1}
\end{figure}

Once source positions were identified, source count rates needed to be estimated. To this aim, it must be recalled that the most precise effective exposure times are evaluated in narrow energy bands. Therefore, we calculated the net (background-subtracted) count rates in four subbands (S1=0.5--1\,keV, S2=1--2\,keV, H1=2--4\,keV, H2=4--8\,keV), and they were later added to form the rates in the broader bands (S, H, and T). We thus first constructed effective exposure maps in these four subbands\footnote{These exposure maps, combined to apply to the S, H, and T bands, were also used in the source detection procedure in a standard way.}. The construction of these exposure  maps assumed a power law spectrum of photon index 1.7 and accounted for the telescope vignetting and bad pixels as well as the quantum efficiency variation of the instrument, including the time-dependent sensitivity degradation, which is particularly important at low energies ($\lesssim 1.5$\,keV). Fig.~\ref{fig:exp_1} shows such a merged effective exposure map, illustrating the features of the bad pixel removal, CCD gaps, observation dithering, etc.\  as well as the overall field coverage. In order to treat uniformly both strong and weak sources, source counts for each subband were then extracted within the 70\% energy-encircled radius (EER) of the PSF, whose size depends on the off-axis angle of the source in the exposure and of the energy band under consideration. A background correction using the background map constructed earlier was applied. Finally, count rates were derived from dividing source net counts by their effective exposure times (values at the source positions in the exposure map of the energy band under consideration), leading to equivalent on-axis values. It should be noted that the presented count rates have thus been corrected for the full PSF and for the effective exposure, which accounts for not only the telescope vignetting, but also the degradation of the detector sensitivity  with time. Therefore, the actual number of counts in a detection aperture is not simply a count rate multiplied by an exposure of 280\,ks. The difference could be up to a factor of $\sim 2$, depending on a source's spectral shape.

We extract an ACIS spectrum for each source detected with $S/N > 10$. 
The on-source spectral extraction circle has the same radius as used for 
the source removal, while the local background spectrum is estimated 
from a concentric annulus with the inner radius equal to 2$\times$ the 
circle radius and the outer radius twice larger. Detected sources are removed from the background 
region. The background spectrum is normalized accounting for bad pixels 
and boundaries of the CCDs as well as the source removal. We obtain the averaged 
response matrices of each source spectrum, using the weights derived from 
on-source 0.5--2\,keV band counts in the detector coordinates of individual 
observations. The spectrum is further adaptively binned to achieve a
background-subtracted signal-to-noise ratio greater than 2.5 in each bin.

We compared the positions of a few well identified X-ray sources with their (known) 
optical counterparts. To this aim, we considered only OB stars, since they are rather 
bright sources of X-rays, whereas no other stellar X-ray emitter was a priori known 
in the field (though several other sources may have possible stellar counterparts, 
see below, but these were not a priori known). We found no significant systematic offset 
($\lesssim$\, 0\farcs5) and therefore, no astrometry correction was applied to the 
X-ray data.  We caution however that these sources lie at large 
off-axis angles, so that the uncertainties in their X-ray positions may be large, but the absence of bright X-ray sources with well-established optical counterparts prohibits us to fine-tune the astronometry to 0\farcs1 accuracy.

\section{Point Source Catalog}

With the detection procedure described in the previous section, we found 165 sources in N11: 43 of them were detected with the highest confidence or smallest $P$ value in the S band, 5 in H band, and 117 in T band. Amongst these, 74 were detected in all three bands, 56 only in two bands, and 35 only in one band (22 for S, 2 for H, 11 for T - while the total band often maximizes the signal-to-noise ratio, some very soft or very hard sources are more easily detected in only the soft or hard band, for example, nearby stars and diffuse emission peaks in the soft band, and faraway accreting sources and AGNs in the hard band). Table \ref{sourcelist} lists for each detected source its position, count rate in the total band and hardness ratios ${\rm HR_1}=({\rm H-S2})/({\rm H+S2})$, and ${\rm HR_2}=({\rm S2-S1})/{\rm S}$, as well as its off-axis angle, number of counts, estimated background counts and {\sl effective} exposure in the detection aperture and in the total band. The last column yields the band where the source was detected with the highest confidence. While count rates may look large at first sight, it must be recalled that the source detection completeness varies across the ACIS-I map,
depending on the local PSF, effective exposure (Fig. \ref{fig:exp_1}), and background. 
Furthermore, count rates have been corrected for CCD degradation, leading even more to apparently high count rates. Without making this correction, the interpretation of the hardness ratios or the count rate to flux conversion given below would depend on the time when observations were taken. The lowest value for source net counts and signal-to-noise ratio are 4.7 and 2.1, respectively.

The conversion from a count rate to an unabsorbed energy flux depends
on the source spectrum and foreground absorption. A characteristic 
value of the conversion is $8 \times 10^{-12}$ 
${\rm~(erg~cm^{-2}~s^{-1}})/({\rm counts~s^{-1}})$ in the 0.5--8\,keV band 
for a power-law spectrum of photon index 2 and an absorbing-gas column 
density $N_{\rm H} \sim 1 \times 10^{21}$\,H-atom\,cm$^{-2}$ (assuming solar 
abundances). This conversion should be a good approximation (within a 
factor of 2) for $N_{\rm H} \lesssim 3 \times 10^{21}$\,H-atom\,cm$^{-2}$. The 
corresponding conversion to a source-frame luminosity in the same band 
is $\sim 2.4 \times  10^{36}$ ${\rm~(erg~s^{-1}})/({\rm counts~s^{-1}})$
at the LMC distance of 50 kpc.

\begin{figure}
\includegraphics[height=5in, clip=]{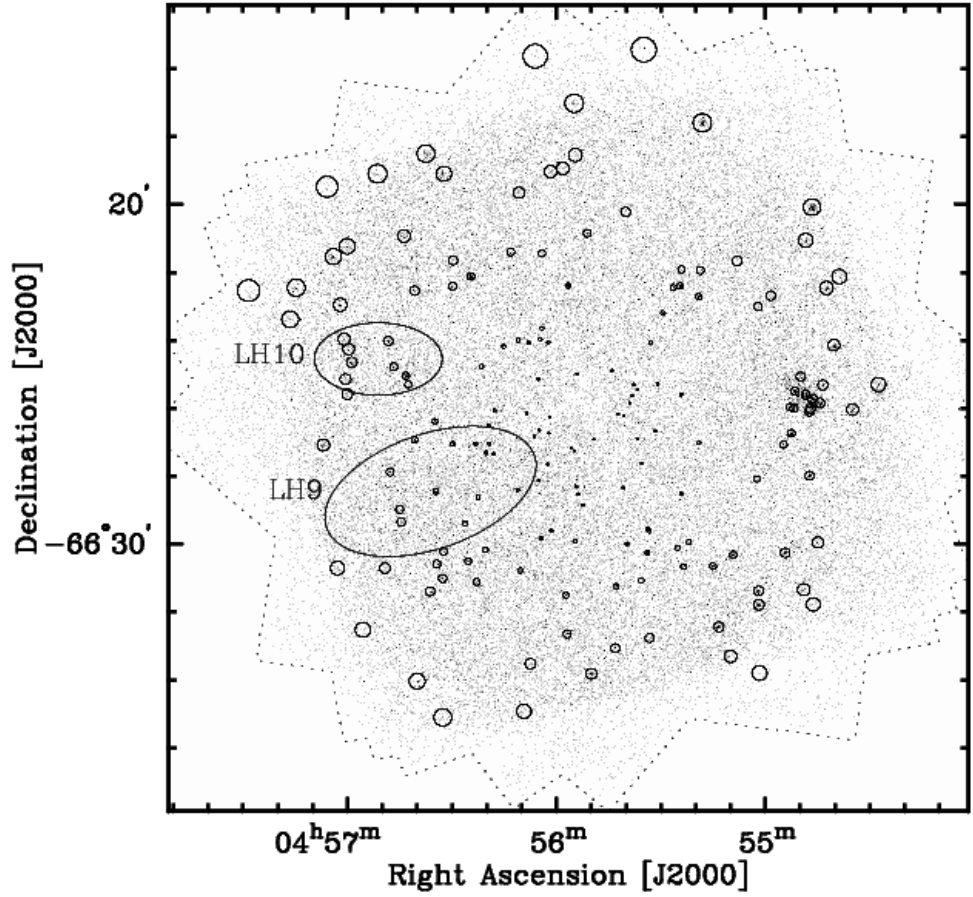}
\caption{{\bf ONLINE MATERIAL} ACIS-I intensity image of \xs\ in the 0.5--2\,keV band and detected sources (Table\,\ref{sourcelist}). The circles mark the regions of individual sources (radius = 1$\sigma$ uncertainty in position, see text for details), the dashed line marks the boundaries of the merged ACIS FoV (Fig.~\ref{fig:exp_1}), and the position of the main clusters are indicated by ellipses.}
\label{fig:x14_sou}
\end{figure}

\begin{figure}
\includegraphics[height=3in]{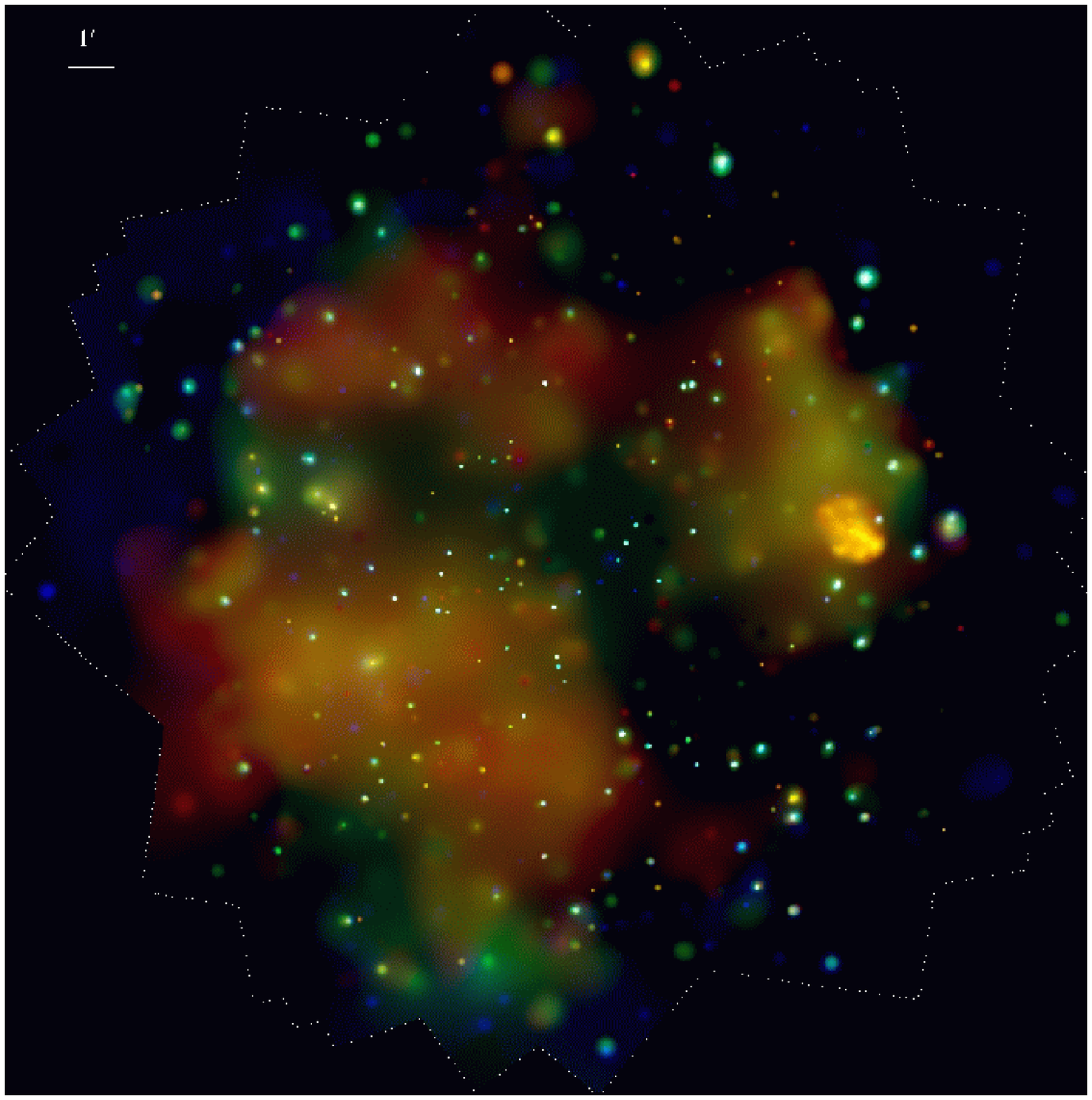}
\includegraphics[height=3in]{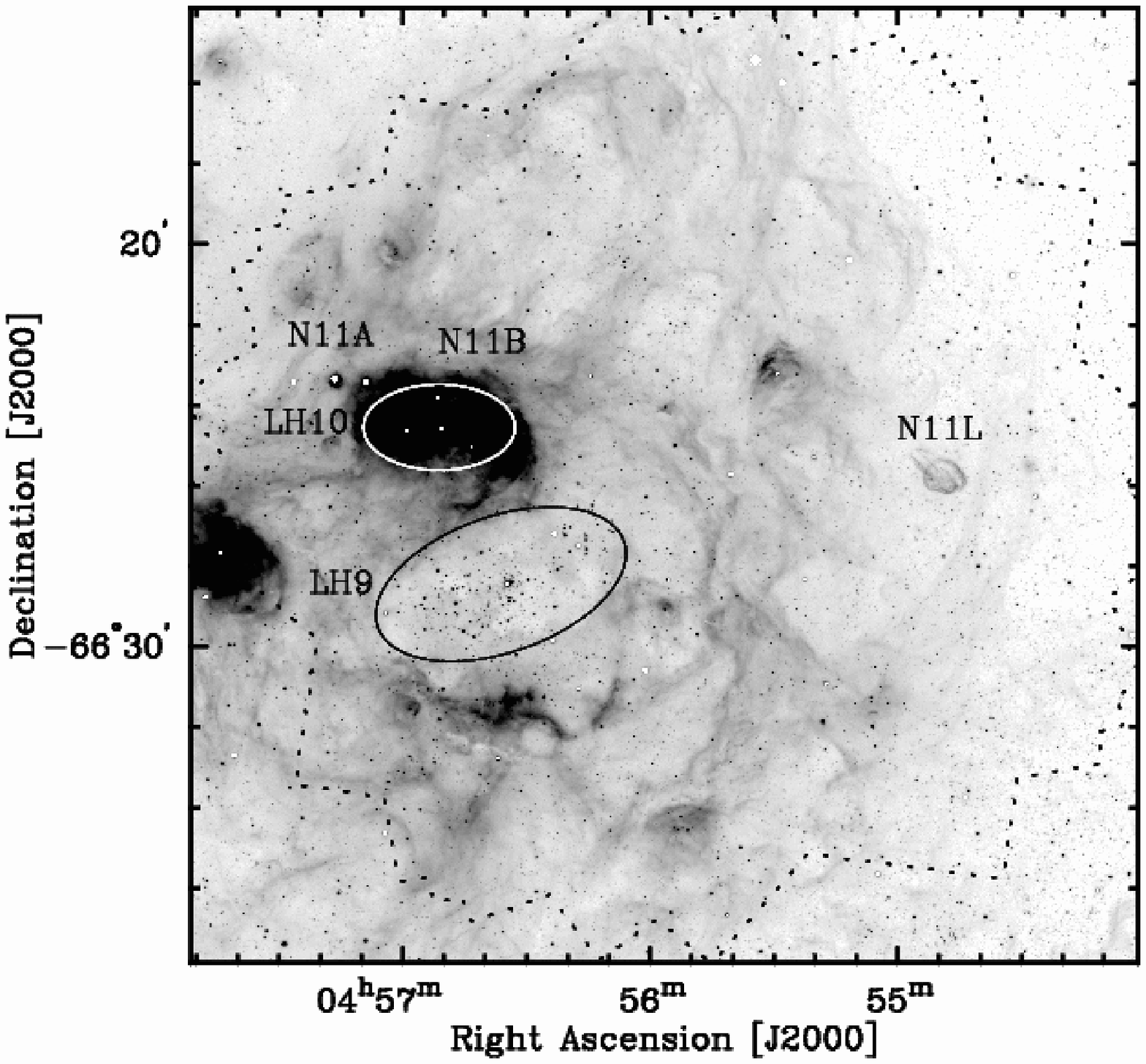}
\caption{{\bf ONLINE MATERIAL} {\it Left:} Tri-color montage of X-ray intensities: (red) 0.5--2\,keV, (green) 2--4\,keV, and (blue) 4--8\,keV. The images are adaptively smoothed with the CIAO routine CSMOOTH. The smoothing scales are calculated separately in the soft and hard bands and with the signal-to-noise ratio $\sim 3$; the subtracted background is estimated locally in CSMOOTH. {\it Right:} H$\alpha$ image for comparison. The main regions are labelled. }
\label{fig:rgb}
\end{figure}

The ACIS-I total band (unsmoothed) image is shown in Fig.~\ref{fig:x14_sou} with 
the detected X-ray sources marked, while Fig.\,\ref{fig:rgb} shows a smoothed 
three-color map of the X-ray emission in N11 as well as an H$\alpha$ image 
for comparison.  The smoothed images show that many point sources are
superposed on diffuse emission.  As the source detection is based on Poisson
statistics and the image smoothing uses Gaussian statistics, there appear to be 
additional point-like sources (to eyes) in the smoothed images.  These are 
artifacts of the smoothing procedure.  We will not consider these false sources
caused by noise bumps.

\begin{figure}
\unitlength1.0cm
\includegraphics[height=1.5in,angle=0,clip=]{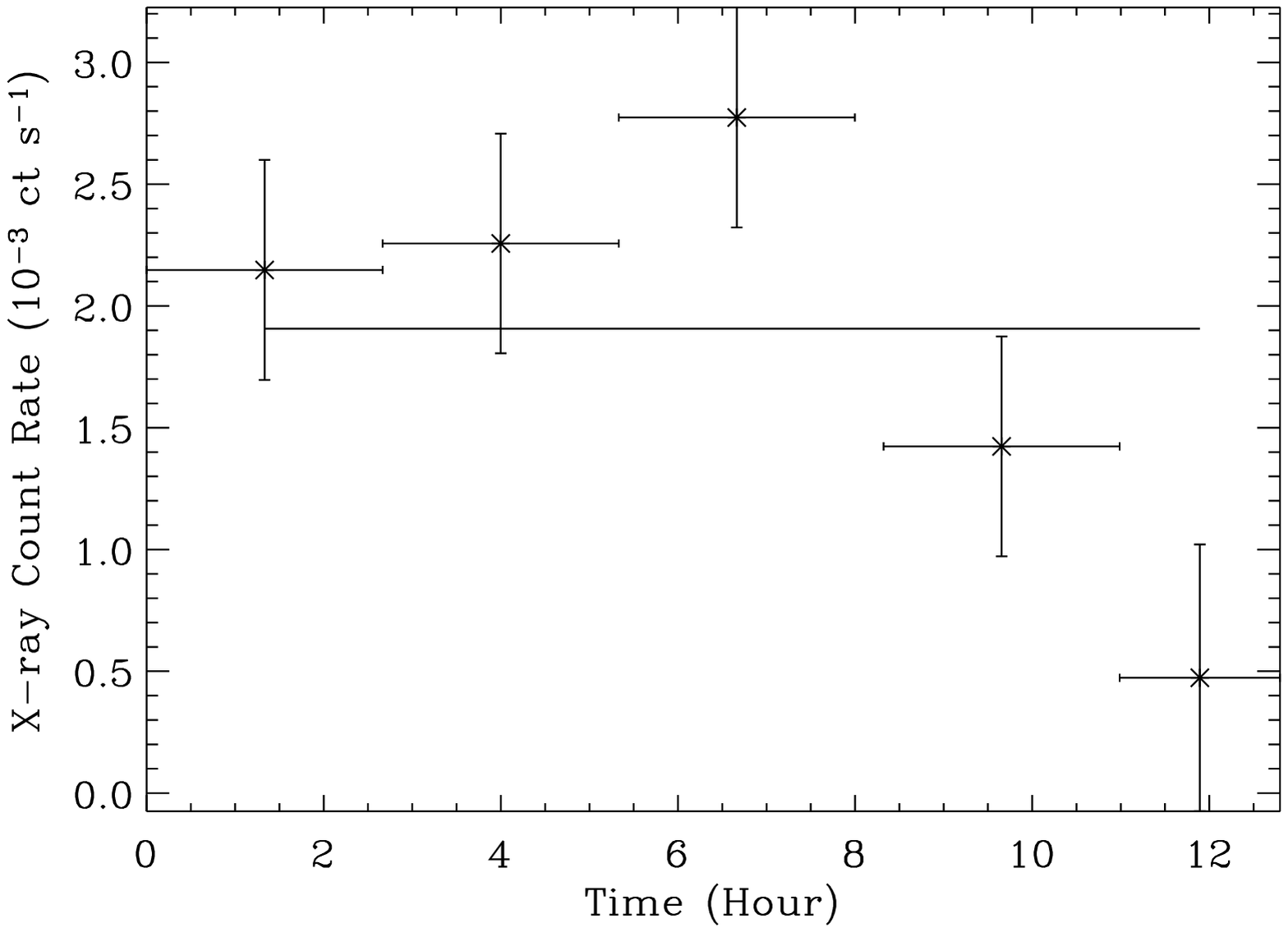}
\includegraphics[height=1.5in,angle=0,clip=]{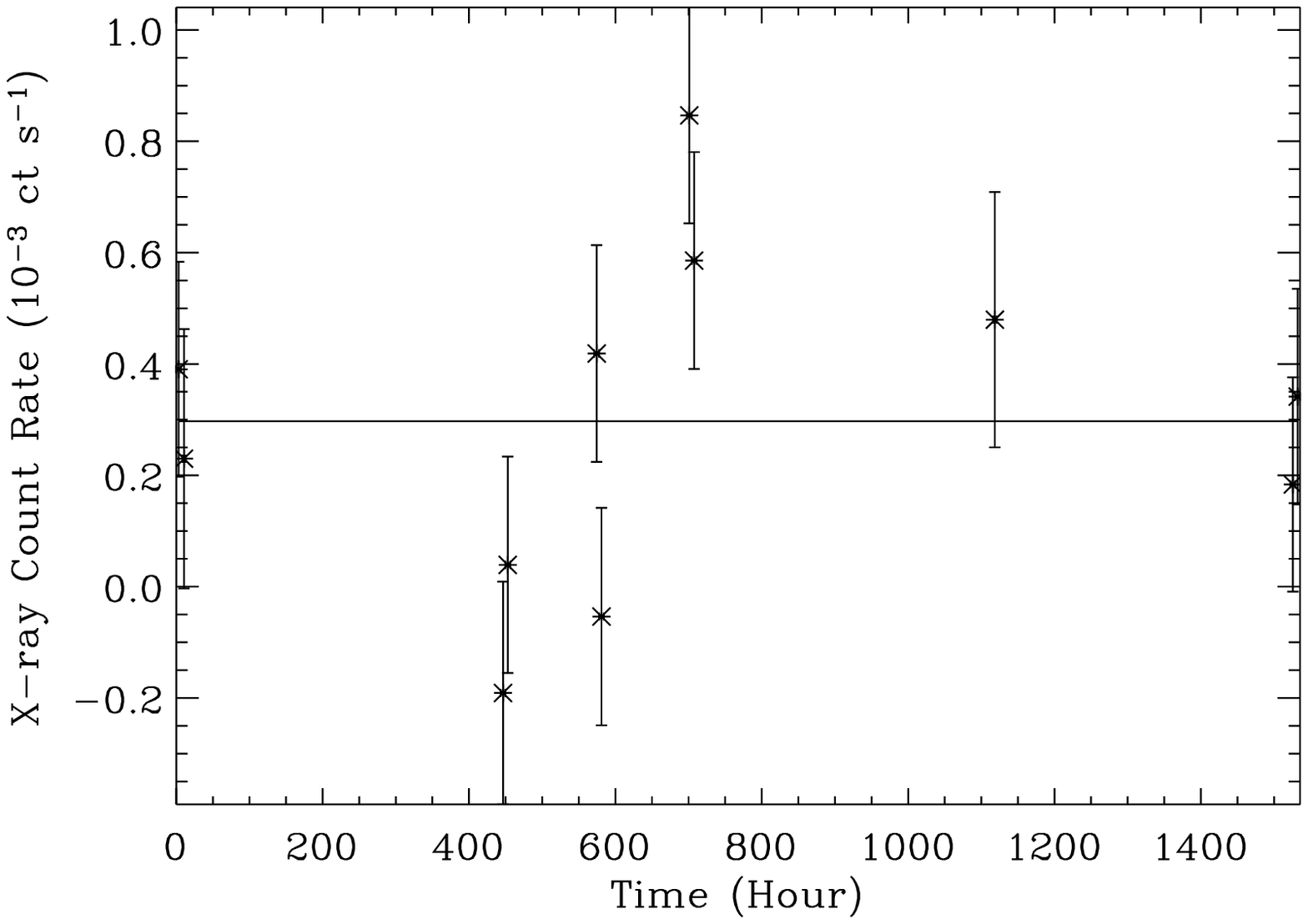}
\includegraphics[height=1.5in,angle=0,clip=]{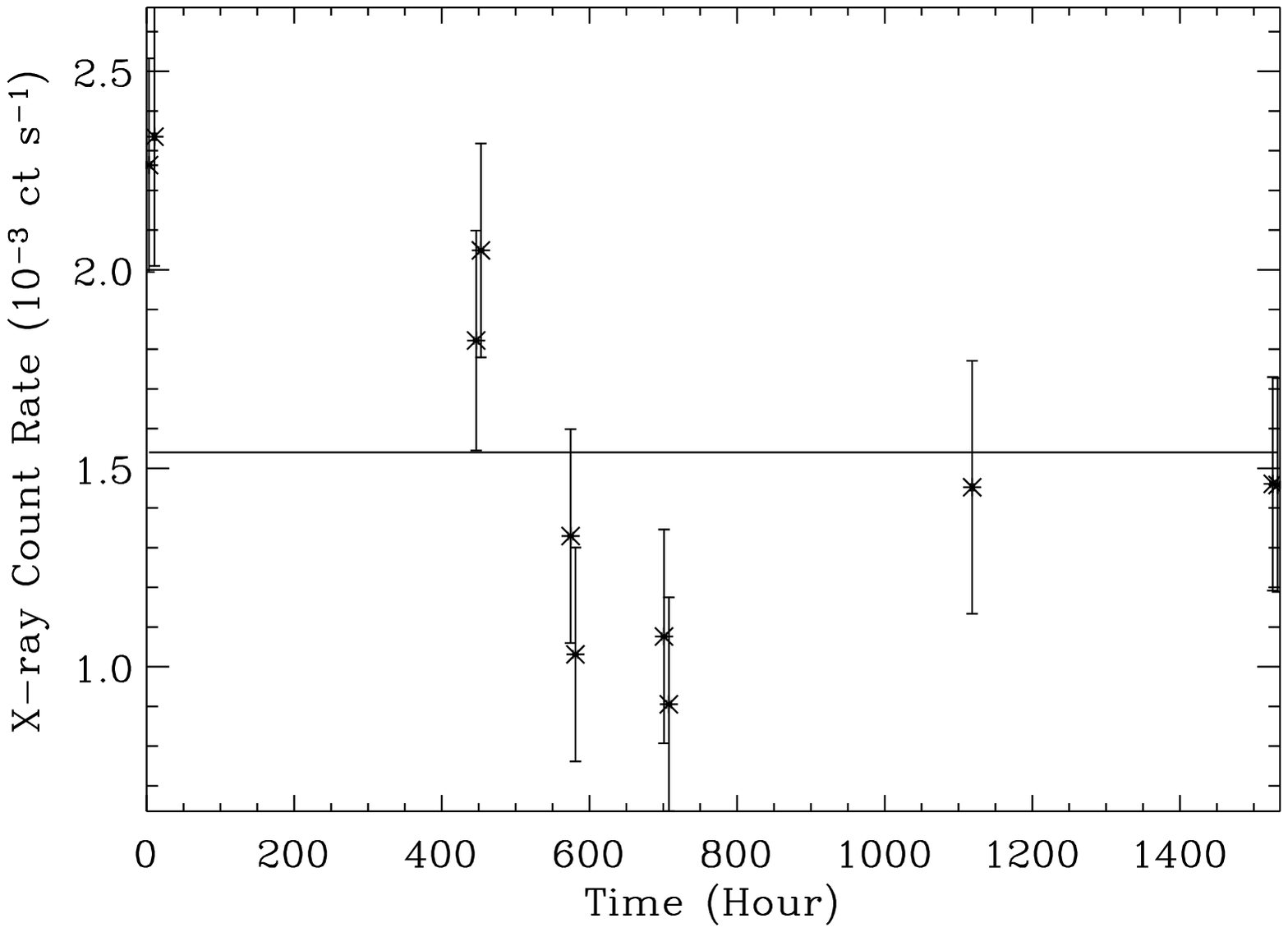}
\caption{Illustrations of temporal variations of three sources:  J045539.69$-$662959.5 (left panel, constancy rejected at 3$\sigma$ level) during one observation; J045509.20$-$663018.5 (middle, constancy rejected at 5$\sigma$ level) and J045702.07$-$662257.1 (right, constancy rejected at 3$\sigma$ level) across the entire data set. The average count rates of individual sources are plotted as the horizontal straight lines. Note that for the last panels, the bins do not correspond to individual exposures but to intervals of time with at least 20 counts. }
\label{fig:timing}
\end{figure}

We also conducted tests for timing variability. We first carried out Kolmogorov-Smirnov tests as well as $\chi^2$ tests on the total band light curves of the 41 sources with detected $S/N > 4$, which are well covered by the six observations. For $\chi^2$ tests, lightcurves were adaptively binned so that each bin contains at least 20 counts. Sources J045509.20$-$663018.5 and J045702.07$-$662257.1 (\# 32 and 158, Fig.~\ref{fig:timing}) show significant variabilities in the total band at confidence levels of 5 and 3$\sigma$, respectively. They possess no counterpart within 1\arcsec\ (see Sect 3.1), and their nature remains unknown. Variability examination in the S and H bands yields no additional results. The analysis of individual observations separately, for all 117 sources that are not near the CCD gaps (with a 12\arcsec\ margin to account for the dithering effect) in any of the observations, yields only one positive result: J045539.69$-$662959.5 (\#64) shows an apparent variation in the S band during the observation \#8210 and at a confidence level of $\sim 3\sigma$ (Fig.~\ref{fig:timing}). This source is a known quasar candidate (see Sect.~3). However, with so many sources studied in two independent bands (the third one being related to the other two since $S+H=T$), this latter variability detection is not inconsistent with the occurrence of such an event by pure chance hence is marginal.

\begin{figure}
\includegraphics[height=3in]{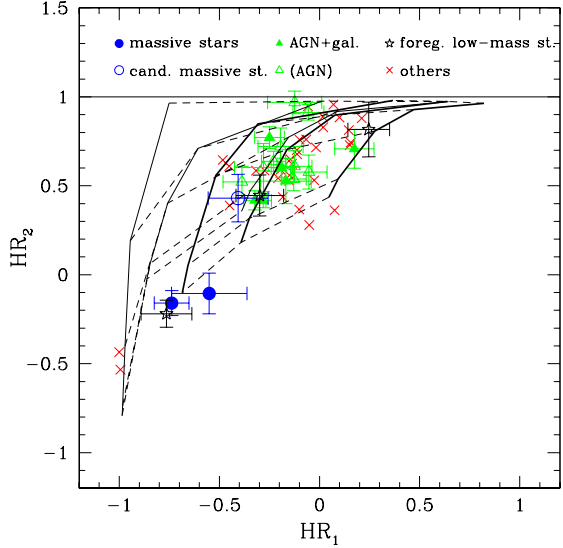}
\caption{Hardness ratios of 53 X-ray sources (error bars, when plotted, represent 1$\sigma$ uncertainties, are $<$0.2, see Table\ref{sourcelist}). Sources are indicated by different symbols according to their nature (see Table \ref{addcount} and top of figure) - note that foreground stars may be active binaries, explaining their high HRs. For clarity, errors are not shown for the sources whose nature is uncertain. Also included in the plot are hardness-ratio models derived from PiMMs: the solid thick curves are for the power-law model with photon index equal to 3, 2, and 1, whereas the solid thin curves are for the thermal plasma with abundance of 0.4 times solar and temperatures equal to 0.3, 1, and 4 keV, from left to right. The equivalent hydrogen absorbing column densities $N_{\rm H}$ are 1, 10, 50, 100, and $300\times 10^{20} {\rm~cm^{-2}}$ (dashed curves from bottom to top). The lowest values correspond to absorptions of sources in the Galaxy or LMC, while the largest values are representative of obscured sources behind LMC. Note that power laws models are just examples, but they may not be representative of the spectral properties of all displayed sources. }
\label{hr}
\end{figure}

\begin{figure}
\includegraphics[width=0.35\textwidth,angle=270.0,clip=]{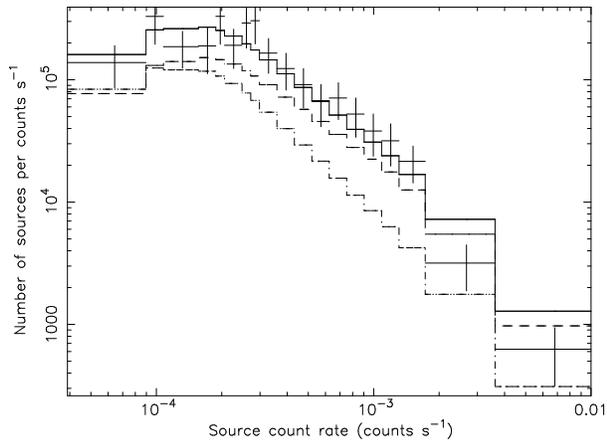}
\caption{Observed differential number-flux relation of the sources best-detected in the 0.5--8\,keV band, compared with the best-fit power-law model. The data are grouped to have a minimum six sources per bin; the fit uses the Cash-statistic and is satisfactory, judged from simulations in XSPEC. The extragalactic AGN component is represented by the dashed histogram, while the power-law component is by the dash-dotted one.}
\label{nfr}
\end{figure}

Figure \ref{hr} shows hardness ratios of the 53 sources with both ratios known with errors less than 0.2: most  (48) sources have $HR_1\sim0$ and $HR_2\sim0.75$, indicating relatively hard sources. To assess the contamination of the catalog by background AGNs, we have characterized the X-ray source number-flux relation (NFR) in N11. This NFR analysis uses only the 141 sources detected in the total band, for homogeneity and to avoid biases towards soft or hard sources. Eddington bias\footnote{The so-called X-ray Eddington bias implies that intrinsically faint sources statistically appear to have higher fluxes than the other way around.} was corrected following the approach of \citet{wan04a}. We may compare the derived NFR to the Log($N$)--log($S$) presented by \citet[and references therein]{mor03}. However, the X-ray absorption through N11 (from NRAO survey\footnote{ http://asc.harvard.edu/toolkit/colden.jsp}, $N_{\rm H} \sim 4.3 \times 10^{20}$\,H-atom\,cm$^{-2}$) is substantially higher than that toward the \chandra\ deep fields (foreground absorption $1.6 \times 10^{20} {\rm~cm^{-2}}$). Correcting for this difference, we find an expected number of AGNs in the field of 91 (Fig.~\ref{nfr}), i.e., most of our sources are in fact extragalactic background objects. Indeed, a few AGNs have been identified in the field: a correlation of our source list with quasar tables in Vizier results in the identification of sources \#11 \citep{kim12,koz09} and \#38, 45, 64, 98, 117, 123 \citep{koz09}. Note however that some parts of the nebula are filled with molecular clouds, and these locally enhanced absorption columns may reduce the number of detectable AGNs.

\subsection{Optical and Infrared Counterparts}

Counterparts to our X-ray sources were searched in several catalogs: USNO-B1.0 Catalog \citep{mon03}, Guide Star Catalog V2.3.2 \citep[GSC,][]{gsc}, 2MASS All-Sky Catalog of Point Sources \citep{cut03}, Magellanic Clouds Photometric Survey \citep[MCPS,][]{mcps}, IRSF Magellanic Clouds Point Source Catalog \citep{irsf}, DENIS Catalogue toward Magellanic Clouds \citep[DCMC,][]{cio00}, and $JHK_s$ photometry of N11 young stellar objects ([HKN2006], \citealt{hat06}).

To find the optimal correlation radius, we first searched for the closest counterpart to each X-ray sources and derived the number of matches as a function of radius: at large radii, the number of matches is proportional to the squared radius, as expected by chance coincidences. A best correlation radius of 1\arcsec\ was found, and used to derive the final list of counterparts (Table \ref{counterparts}): 71 of the 165 sources have at least one counterpart within 1\arcsec. Amongst these, thirteen objects are known massive stars (see Table \ref{detO} and Sect. 4), two are OB candidates (\#130 and 136, \citealt{hat06}), and one is a HAeBe candidate (\#63, \citealt{hat06}). Two additional sources have been misidentified with stars in the past: Src \#98, proposed to be a HAeBe candidate on the basis of the photometry \citep{hat06}, is in fact a quasar \citep{koz09}, while Src \#24, identified to be a young stellar object \citep{whi08}, actually corresponds to the nucleus of a background galaxy \citep{gru09}.

\begin{figure}
\includegraphics[height=3in]{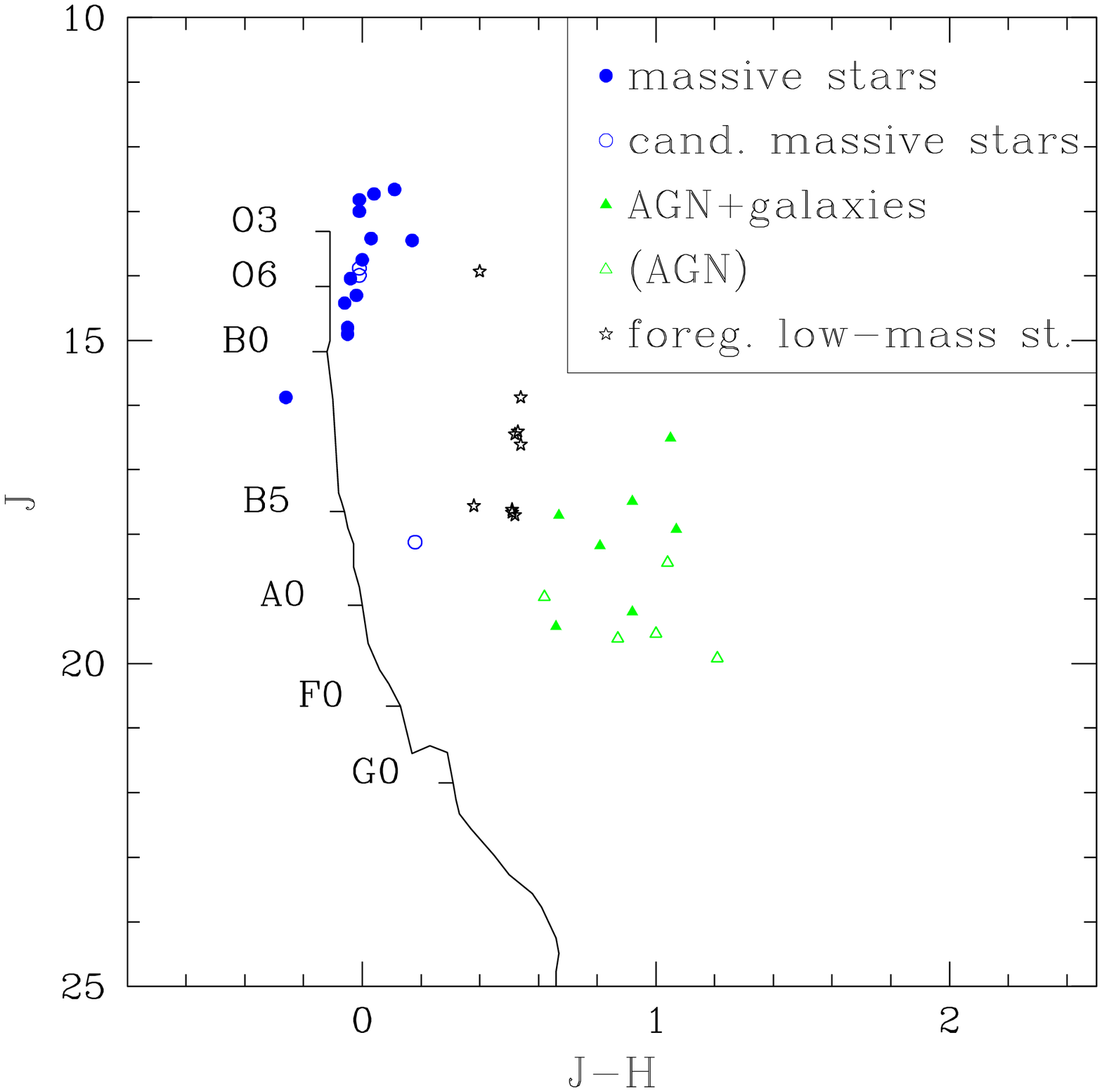}
\includegraphics[height=3in]{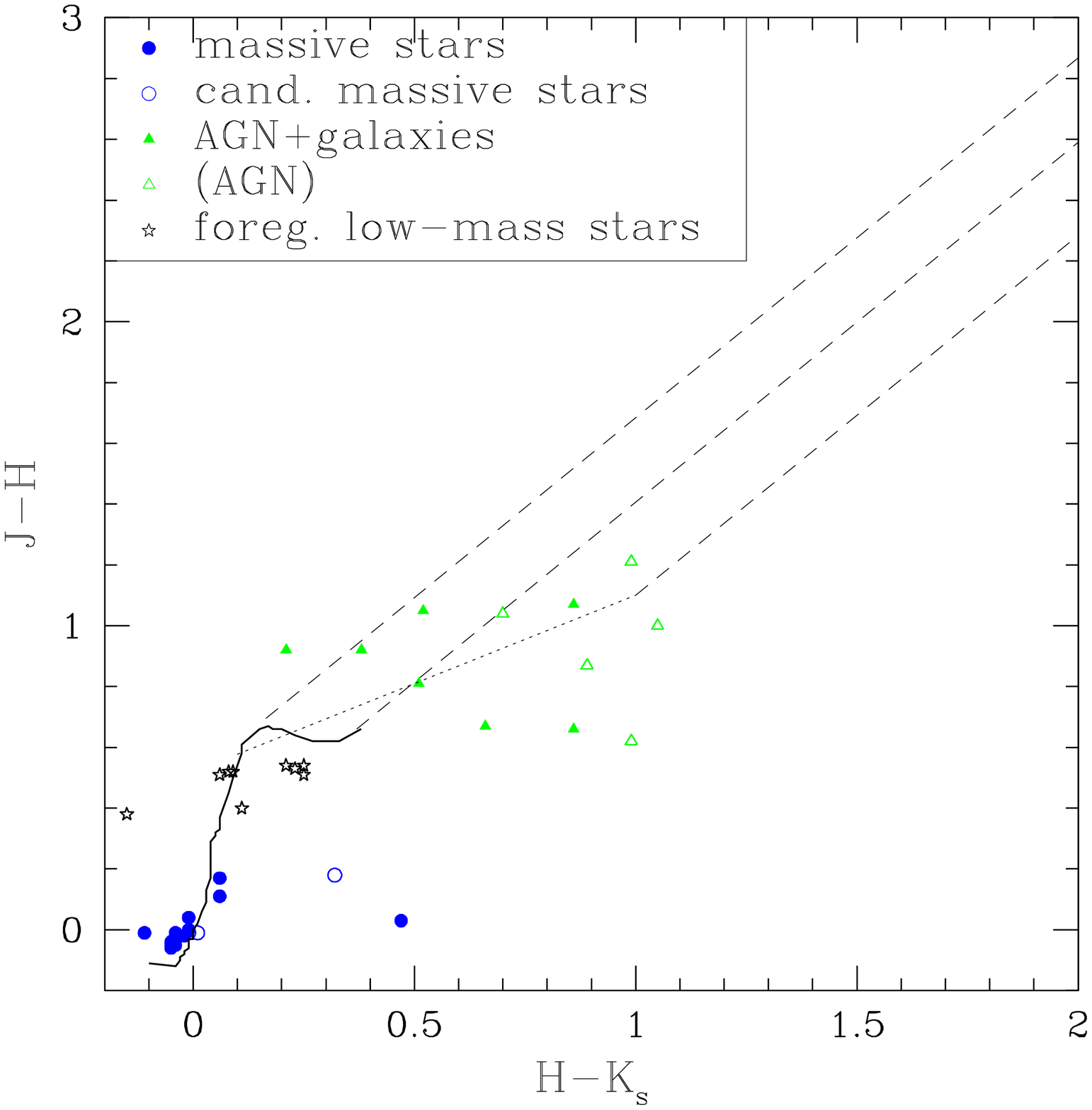}
\caption{Color-magnitude (left) and color-color (right) diagrams for the 37 X-ray sources with IRSF counteparts with full $JHK_s$ photometry. Sources are indicated by different symbols according to their nature (see Table \ref{addcount}). The main sequence magnitudes and colors are taken from \citet{mar06} for O-stars (masses $>15$\,M$_{\odot}$) and \citet{tok00} for the other spectral types (masses in the range 0.1--15\,M$_{\odot}$); it was shifted by $DM=18.5$ for the color-magnitude diagram. The dotted line shows the intrinsic (i.e., dereddened) colors of classical T Tauri stars \citep{mey97}, the dashed lines correspond to increasing values of absorption (using $R_V=3.3$ and \citet{car89} extinction law) for the blue and red limits of the main sequence for low-mass stars and of the TTs sequence.  \label{HR}}
\end{figure}

The photometric measurements of the counterparts appear coherent in different catalogs. We therefore focus on IRSF, because it contains the largest number of counterparts (58 sources) amongst the tested catalogs. Considering the sources with full $JHK_s$ photometry available, color-magnitude and color-color diagrams can be constructed (Fig.~\ref{HR}). Besides massive stars, counterparts appear to the right of the main sequence, suggesting that they are young, still forming stars; however, known quasars also have similar photometric properties, requiring additional investigation.

To this aim, we further used H$\alpha$, [\ion{O}{3}], and [\ion{S}{2}] images taken with the MOSAIC camera on the Blanco 4m telescope at the Cerro Tololo Inter-American Observatory.  The H$\alpha$ observations consisted of three dithered exposures of 300\,s each for each location; the bulk of N11 was imaged on 2008 December 5 and the periphery of N11 was imaged on 2010 January 9.  The [\ion{O}{3}] $\lambda$5007 observations and the [\ion{S}{2}] $\lambda\lambda$6716,6731 observations consist of four dithered 450\,s exposures for each location; these images were obtained on 2011 October 31.  We have also used infrared observations made with the \emph{Spitzer Space Telescope}.  The \emph{Spitzer} images and photometry of point sources of N11 are taken from the previous work by \citet{gru09}, who made a photometric catalog and identified young stellar objects for the entire LMC. These images have been used for inspecting the counterparts to the X-ray point sources.  Among the three optical images, the [\ion{S}{2}] image is the most useful because the diffuse emission from ionized gas is not as strong and confusing as that in H$\alpha$ and [\ion{O}{3}]. Among the infrared images, we have used primarily the 3.6 and 8.0 $\mu$m images.

Figure \ref{map} shows $10'' \times 10''$ cutout [\ion{S}{2}] images 
overplotted with error circles (radius from Column 3 of Table 2) centered 
on the derived positions of all 165 X-ray sources.  These images are useful 
for an independent confirmation or rejection of optical counterparts. 
Although some of the error circles have radii $\ll 1''$, the combined errors
in the X-ray and optical astrometry may reach $\sim 1''$.  We have 
conservatively use a minimum correlation radius of 1$''$ for identification
of optical counterparts; however, only 3 sources (\#43, 78, and 107) have
optical counterparts that are within 1$''$ but outside the error radius.
The results of our investigation are noted in Table \ref{addcount}, 
including \emph{Spitzer} counterparts of the X-ray point sources.

To assess the physical nature of the optical and infrared counterparts
of the X-ray sources, we have assembled spectral energy distributions
(SEDs) for the sources that have photometric measurements available.
The following passbands and catalogs have been used: $UBVI$ from MCPS,
$JHK_s$ from 2MASS and IRSF, and IRAC bands from \citet{gru09}.
These SEDs are presented in Figure \ref{sed}.

The SEDs of known massive stars, such as sources \#48, 77, 97, 110, 115, 
126, 141, 142, 146, and 157, have a distinct shape that falls off toward 
long wavelengths, following the Rayleigh-Jean Law.  
While the SEDs of stars can be diagnosed by their downturn in infrared, the nature of the stars need to be estimated from their photometric colors and magnitudes.  For each object with a stellar SED, we use the MCPS $UBVI$ and IRSF $JHK_{\rm s}$ photometry to determine colors in several combinations of bands (such as $U-B$, $B-V$, $V-I$, $J-K$, etc.), compare the observed colors with those of dwarfs (luminosity class V) and supergiants (luminosity class I) to assess its spectral type, and compare the observed magnitude with the expected absolute magnitude to determine its distance.  

We find that Sources \#130 and 136 are B2 giants, in agreement with the suggestion of candidate OB stars by Hatano et al. (2006), sources \#108, 139, and 145 are late-type B dwarfs, and source \#106 may be of type A0.  Distances that we derived for these objects are $\ge$50 kpc, and they are thus in the LMC. Using the same conversion factors as for known massive stars (see Sect. 4.2), the X-ray luminosities of these sources, in the total band, amount to $\sim5\times10^{32}$\,erg\,s$^{-1}$ for the three late-B and the A0 star, and $1-2\times10^{33}$\,erg\,s$^{-1}$ for the two B2III stars. This leads to \loglxlbol\ of $-2.6$ for the A0 star, $-3.2$ for the late-B stars, and $-4.6$ for the B2III stars. Such luminosities are too high for even flares of PMS companions or flares of these stars themselves \citep{rob11}, but too low in comparison with those of known HMXBs in the LMC \citep[$>10^{34}$\,erg\,s$^{-1}$, ][]{sht05}. Furthermore, such high \loglxlbol\ ratios are not compatible with embedded wind shocks (such stars would not have the strong mass-loss necessary for this mechanism, anyway). In the absence of further information, it seems likely that objects other than these AB stars are the true X-ray emitters: chance alignment with foreground (soft) source or background (hard) source or localized peak in soft diffuse emission. In this context, it should be noted that \# 106, 108 and 130 seem to emit mostly soft X-rays, while the emissions of \#136 and 139 appear much harder.

The cooler stars are all in the Galactic foreground: source \#56 is a known K0 star; sources \#5 and 28 are G stars; 25, 43, 78, 87, and 109 are K stars; 54, 95, and 107 are M stars.  
In two cases, sources \#25 and 109, the colors at long wavelengths suggest K spectral type, but the star appears too bright in $UB$.  It is likely that they are Sirius-like systems in which the K star dominates the emission in longer wavelength and the white dwarf dominates the emission in shorter wavelengths.  The individual results of the stellar counterpart analysis are given in Table \ref{addcount} and a summary by category is provided in Table \ref{countsum}. 

AGNs and galaxies can also be diagnosed from the SEDs by their distinct
shape that rises toward long wavelengths in the infrared \citep{don08,dey08}.
For example, sources \# 11, 24, 36, 38, 45, 63, 64, 98, 117, and 123 have well populated SEDs rising in the IR and are thus good candidates for AGNs. Indeed, seven of them (\#11, 38, 45, 64, 98, 117, and 123) have been identified to be quasars by \citet{koz09} and one (\#24) is a resolved galaxy with a prominent nucleus \citep{gru09}.  Since stellar emission is not expected to be flat and since many of these objects are detected only in the IR (not in optical), 
we suggest that objects with flat SEDs may also be AGNs. In Table \ref{addcount}, we use ``AGN''  to denote confirmed objects and ``(AGN)'' for candidates.  

Finally, our observation did not reveal active HMXBs amongst known massive stars in N11 (see Sect.\ 4) and the late B-stars newly identified with the photometry (see previous paragraphs) display no definitive sign of being in HMXBs. We have further examined the data to see whether other bright X-ray sources would have (unknown) OB counterparts, following the method of \citet{sht05} and using the information available on the nature of the sources, when existing.  As in 30 Dor \citep{tow06b} and some other LMC clusters \citep{osk05}, no HMXB was detected in N11, although its cluster LH9 is old enough ($7.0\pm 1.0$\,Myr, \citealt{mok07}) for its initially most massive stars to have undergone supernovae. Adopting the classical Salpeter initial mass function (IMF) for the observed stellar content and with upper cut-off at $150\,M_\odot$, up to 10 stars initially more massive than $60\,M_\odot$ might have been present in the cluster. However, the production efficiency of HMXBs is low, which could explain the absence of detections in N11 \citep[e.g.,][though not necessarily in agreement with Galactic census of HMXBs from \citealt{hel01}]{osk05,cla08}. 

\section{Massive Stars}

\subsection{The Massive Star Population in the {\em Chandra} Field}

Of the four massive star clusters in N11, two (LH9 and 10) are covered by our {\em Chandra} observation. 
Their stellar content is quite well known. The first complete study \citep[][hereafter PGMW]{par92} provided 
spectroscopic classification for about 75 stars, among which 43 are O-type 
objects. Using {\em HST} data to disentangle the two 
compact OB groups HD32228 in LH9 and PGMW3204/9 in LH10, \citet{wal99} gave spectral classification for 
20 additional hot stars, while the VLT Flames Survey led to the discovery of 25 additional O-type stars,
 including a potential runaway star of type O2.5III (\citealt{eva06}, hereafter ELST\footnote{Note that 
 4 O and 9 B of ELST stars are outside the \chandra\ FOV.}).
In total, 1 Wolf-Rayet, 81 O-type stars and 80 B-type stars are now known 
in the \chandra\ FOV. The spectral monitoring of ELST further indicated 
a binary fraction of 36\% amongst massive stars of LH9-10, which is quite low for such stars \citep{san10}. Finally, \citet{mok07} performed 
atmosphere modelling on 22 of the ELST targets, leading to 
the first accurate determination of their physical properties. 

These studies showed that LH9 and LH10 appear very different from each other. 
Indeed, LH9 is the most extended and the 
richest cluster in N11. It is dominated by a compact group of stars collectively referred to as HD32228, which contains one Wolf-Rayet star of the carbon sequence (Brey 9) and many late O-type stars \citep{wal99}. Furthermore, stars in LH9 have blown a large superbubble \citep{ros96}. In contrast, the LH10 cluster, 
still partly embedded in its natal cloud, appears rather young, with wind-blown bubbles of limited size \citep{naz01} and earliest O-type stars still present \citep{eva06}. The IMF in these clusters was constrained to 
$\Gamma_{\rm LH10}=-1.1\pm 0.1$ and $\Gamma_{\rm LH9}=-1.6\pm0.1$ (PGMW),
confirming that there are many more high-mass stars in LH10 than in LH9. 
These facts, together with the higher reddening of LH10, indicate the
relative youth of LH10 compared to LH9 (about 2\,Myr difference, see PGMW and 
\citealt{mok07}). Because of the age difference and of the position of LH10 at the periphery
of the LH9 superbubble, it was suggested that the pair constitutes
an example of sequential star formation \citep{wal92}.

\subsection{Detection of Massive X-ray Emitters in N11}

The poorer PSFs of the previous X-ray observations of N11 
({\em ROSAT} - \citealt{mac98,dun01}; {\em XMM-Newton} - \citealt{naz04}) 
did not lead to unambiguous detections of X-ray emission associated 
with the OB stars of LH9 or LH10. At first sight (see Fig.~\ref{fig:x14_sou}), 
the {\em Chandra} X-ray point sources do not appear obviously clustered 
towards LH9, LH10 or their periphery, due to the large number
of background sources (Sect. 3). However, correlating the list of X-ray 
sources (see Table\,\ref{sourcelist}) with the list of OB stars with known spectroscopic classification
led to 13 positive matches. As a last check, we further inspected the X-ray image by eye and one additional X-ray source associated with an O-type star was clearly spotted (see e.g. the many countours showing a source associated with PGMW3100 in Fig. \ref{lh10hst}). An extraction run specific to OB stars (see below) yields a 2$\sigma$ detection for this object and a false detection probability just below that of the detected objects (i.e. much higher than for the truly undetected O-stars).   It lies near a rather bright point source and on the edge of 
diffuse emission, which probably explains its non-detection by the automated 
detection algorithms. 

Table\,\ref{detO} lists the properties of these 14 OB stars with detected 
X-ray emission. The first five columns report the X-ray source number, 
the ELST or PGMW identification, the spectral type and binary status 
(from ELST, \citealt{wal99}, or PGMW). 
In addition, we derived for each object the color excesses $E(B-V)$ from 
the $BV$ photometry, using the intrinsic colors from \citet{mar06}. Galactic 
reddening towards the LMC was estimated to be 0--0.15 mag by \citet{ost95}; 
as the reddening of some of our stars is $\sim0$, we will consider the 
Galactic contribution negligible and we therefore calculate the absorbing 
columns $N_{\rm H}$ using a gas-to-dust ratio of 
$N_{\rm H}/E(B-V)=2.4\times10^{22}$\,H-atom\,cm$^{-2}$\,mag$^{-1}$, 
typical of the LMC \citep{fit86}. The bolometric luminosities, when not 
estimated by \citet{mok07}, were also derived from the $BV$ photometry with 
intrinsic colors and bolometric corrections from \citet{mar06}; they were 
derived using the properties of the earliest component if the object is 
a binary with types of the two components known (`SB2' in Table\,\ref{detO}).  
These derived absorbing columns and bolometric luminosities are 
presented in the sixth and seventh columns of Table\,\ref{detO}. The last 
two columns of that table give the unabsorbed flux (in the 0.5--10\,keV 
band) and \loglxlbol\ ratio (see also Fig. \ref{lxlbol}). These X-ray 
luminosities were derived by different methods depending on the source 
brightness. Five of the sources have enough counts ($>$80 cts) for a rough 
spectral analysis. They were modeled within XSPEC using an absorbed thermal 
model (XSPEC models mekal and vphabs, both with metal abundances set to 
0.3 times solar). Results of these fits are shown in Table\,\ref{specO}. 
As is usual for massive stars \citep[e.g.,][]{naz10}, two solutions of rather similar quality can be found, one with high T and low $N^{add}_{\rm H}$, the other with low T and high $N^{add}_{\rm H}$. It is difficult to choose between the two solutions since the exact amount of additional absorption $N^{add}_{\rm H}$, due to the cool wind, beyond the interstellar absorption cannot be a priori fixed: the formal best-fit one is listed in Table\,\ref{specO}. However, it must be stressed that this dual solution ambiguity affects only the derivation of intrinsic emission levels, but has no impact on the derivation of \lxlbol\ ratios, as the X-ray luminosities used in this context are corrected by the {\em interstellar} absorption only. For the fainter sources, we derived fluxes by converting the 
count rates. Using the on-axis response matrices and a thermal model 
with a temperature $kT$=0.6\,keV absorbed by the $N_{\rm H}$ determined 
above (both with metal abundances set to 0.3 times solar), we derived 
conversion factors between count rates (in cts~ks$^{-1}$) and unabsorbed 
fluxes (in erg\,cm$^{-2}$\,s$^{-1}$) of 0.5--2$\times10^{-14}$. Using a 
temperature of 0.3\,keV would increase the fluxes by less than a factor of 2.

Only one of the detected sources is a B-type star, the detection fraction 
for these objects is thus low, $\sim$1.3\% (1 out of 80). This is not surprising 
since B-type stars are generally not bright X-ray emitters and the X-ray emission 
apparently associated with such objects has generally been attributed to a 
flaring PMS companion \citep[see e.g.,][]{san06,eva11} or to an accreting compact 
companion. Generally, the B stars with detected X-ray emission have early 
types (B0-B0.5) and their emission is due to embedded wind-shocks as in O-type stars. This could apply to Src \#110 (ELST33), an early-type B star whose count rate is correspondingly low with a signal-to-noise ratio of $\sim2$ and whose \lxlbol\ is similar to that of O-stars in our sample.

Most of the detected sources correspond to hot, O-type objects. Two of 
them correspond to the compact OB groups HD32228 and PGMW3204/9, while 
another one is associated with the compact \ion{H}{2} region N11A and its small 
cluster (see individual discussion for these below). Excluding the 19 stars 
in these compact groups, the fraction of detected O-type stars is 16\% (10 out of 62). When 
ranked by magnitude or spectral types, it appears that the hottest and/or 
earliest objects are preferentially detected, as could be expected 
(the detection fraction is 57\% for O2--5 stars, 25\% for 
O5.5--O7 stars and 3\% for O7.5--9.7 stars). However, when one looks
into details, things are not so simple. For example, of the four O2--3 stars 
present in the field, only two are detected: the one belonging to the 
PGMW3204/9 compound (which thus contains additional massive stars emitting X-rays) and ELST31. The latter object is neither a known binary nor the brightest earliest-type star (ELST26 being slightly brighter than ELST31). 

To gain further insight into the global properties of the OB star population, 
we estimated the total-band count rates of all OB sources in the field using their 
catalogued positions and corresponding 90\% EER (i.e. not 70\% EER as before) to minimize the potential effect of the astrometry errors in the X-ray data and the objects' positions. Results for the 13 detected O-stars are of course consistent with the count rates reported in Table \ref{sourcelist}. On the other hand, for the 52 undetected O-stars, these measurements allowed us to derive upper limits, which we adopted to be the 3$\sigma$ errors on the count rates. These upper limits on the count rates were transformed 
into upper limits on the unabsorbed flux as was done above for the fainter 
detected sources. 

Table\,\ref{allO} summarizes the properties of these 52 undetected O-type 
stars present in the \chandra\ FOV. Columns one and two give the star 
numbers in ELST and PGMW, respectively; the third and fourth columns 
indicate the spectral type and binary status; the fifth column provides 
the bolometric luminosity as given by \citet{mok07} when available or 
as derived from the $BV$ photometry otherwise; the sixth column lists the 
absorbing columns, derived from the photometric color excess as explained earlier; the 
last columns yield the upper limits on the X-ray luminosity and on the 
\loglxlbol\ ratio (see also Fig.\,\ref{lxlbol}). In addition to these limits, we also derived the combined properties of these 52 undetected O-stars. Summing their count rates and combining the associated errors yields a stacked value of $0.7\pm0.4$\,cts\,ks$^{-1}$, corresponding to an X-ray luminosity of $(2\pm1)\times10^{33}$\,erg\,s$^{-1}$ and a \loglxlbol\ of $-7.3\pm0.3$. This value is expectedly lower than all individual lower limits on \loglxlbol. It also implies that several undetected massive stars (since the stacked signal appears dominated by $<$10 objects) emit with levels not much below our sensitivity limit, but only more sensitive observations with a high spatial resolution will be able to clearly detect them above the bright diffuse background.

Having all that information at hand, we may now compare N11 with the Galaxy. 
In our Galaxy, massive O-stars display soft spectra ($kT$ of 0.2--0.6\,keV) and follow $L_{\rm X}\sim 10^{-7}\times L_{\rm BOL}$ \citep{har79,lon80,pal81,sci90,ber96,naz09,naz11}. This relation reflects an intimate link of X-rays and the stellar winds, as X-rays are produced behind shocks in these unstable outflows, though a full understanding of the origin of the relation is still lacking \citep[steps in this direction have been made by][]{osk11,owo}. Different \loglxlbol\ values have been reported in the literature for Galactic objects, but these differences are probably explained by choices made in the analysis (e.g., method for deriving $L_{\rm BOL}$) and the data quality \citep{naz11,naz13}. Since the X-ray emission depends on the winds, it may be expected that massive stars with different wind properties, such as low-metallicity, will show a different level of X-ray emission. In addition, deviations from that `canonical' $L_{\rm X}-L_{\rm BOL}$ relation are also found in exceptional cases. For example, while most massive O+OB binaries are not much harder or brighter (small overluminosities of 0.2dex at most, \citealt{osk05,san06,naz09,naz11,naz13}) than single stars, a few systems appear hard and overluminous because an X-ray-bright wind-wind collision is present (e.g., HD93403, \citealt{rau02}; Cyg\,OB2\#9, \citealt{naz12}). Magnetically confined winds may also lead to overluminosities in strongly magnetic stars (e.g., $\theta^1$\,Ori\,C, \citealt{sch00,gag05}). 

In N11, results from spectral fits suggest rather large \loglxlbol\ ratios  and high plasma temperatures. Indeed, except 
for the HD32228 compound, all detected sources clearly lie above the 
Galactic \loglxlbol$=-7$ relation (Fig.\,\ref{lxlbol}) 
and temperatures of 0.7--1.3\,keV are recorded here, when temperatures of 0.2 or 0.6\,keV (depending of the trade-off between temperature and absorption mentioned above) are usually observed for Galactic O-stars. While it is possible that the stellar census is limited by confusion (with several neighbouring stars mistaken as one, leading to errors in the bolometric luminosity estimates), it is also very probable that we detect only the X-ray-bright tip of the massive star population. The detected objects would then be wind-wind interacting systems or strongly 
magnetic objects; they would therefore not be fully representative of the 
properties of the O-type population in N11. In this context, it may be worth noting that, excluding the stars in compact groups, only 20\% (2 out of 10) of the X-ray sources are known binaries, i.e., a smaller fraction than in the N11 population (36\%, ELST) - though the limited monitoring of ELST may have missed the multiplicity of some objects. Further study should be 
undertaken to clarify the status of the detected objects (bona-fide single 
stars, colliding-wind binaries, magnetic objects). To get a more representative idea of the actual X-ray emission level from massive stars in the LMC, we may turn to the undetected O-stars: their stacked emission suggests a lower \loglxlbol\ value of $-7.3$. This is a value comparable to Galactic values (e.g., Carina nebula, \citealt{naz11}), which contradicts a priori intuitions (lower metallicity$\rightarrow$weaker winds$\rightarrow$fainter X-ray emission). Future observations are however needed to confirm the overall representativity of this \loglxlbol\ value for massive stars, and to enlarge the study of the \lxlbol\ ratio in the LMC.

\subsubsection{Sources in LH9: HD32228 and Its Environment}

The main component of LH9, the compact group also known as HD32228, is 
clearly detected in our observations (Fig. \ref{lh9ha}). Its X-ray emission 
follows the visible one, i.e., it is not a simple point source and looks rather
 extended. The morphology is actually reminiscent of a point-like source 
superimposed on a small region of diffuse emission, itself immersed in 
the fainter superbubble emission from the whole cluster. We have extracted 
spectra of each of these 3 regions (see Tables \ref{specO} and \ref{specdif}). 

The emission peaks at the position of the stellar group encompassing the Wolf-Rayet  Brey 9 or 
BAT99-10. 
Since this evolved object belongs to the carbon sequence, it is not 
expected to emit significant amounts of X-rays if single \citep{osk03}. The 
total luminosity from all O-type stars in the HD32228 compound amounts to 
$\sim7.5\times10^{39}$erg\,s$^{-1}$, implying a \loglxlbol\ ratio of 
$-7.1\pm0.1$, fully compatible with the Galactic value (Table \ref{detO}). 

The surrounding area (`near HD32228', see Table\,\ref{specdif}), once 
cleaned of the superbubble `background' contribution, presents a harder 
spectrum than the emission associated with the superbubble itself. Its high 
temperature and high absorption suggests the emission to be dominated by 
the overall X-rays from unresolved stellar objects, most probably O and PMS stars. 
In contrast, another region in LH9 with higher surface brightness 
(`inside LH9', see Table\,\ref{specdif}) displays a spectrum very similar 
to the superbubble itself - only a few deviant high-energy bins, maybe 
spurious, are detected as a very high temperature plasma with low emission measure. To 
the limits of our data, the X-ray emission associated with the superbubble around LH9 thus appears 
rather uniform if one excludes HD32228 and its surroundings.

\subsubsection{Sources in LH10}

Fig. \ref{lh10hst} shows a close-up on the LH10 cluster. In this region, 
six point-like sources are detected. They correspond to five O-type stars 
(ELST 31, 38, and 50, PGMW 3070 and 3120) and one compact group 
(PGMW3204/9). The brightest source, PGMW3070, appears actually 
as a tight cluster in the {\em HST} images. The point-like source associated with 
the two subclusters display a slightly elongated shape, probably 
indicating that they are not truly point-like sources. Their flux is also too 
high for a single star, with \loglxlbol\ of $-6.0$ for the sole 
PGMW3070. The fluxes of the other sources in LH10 are also higher than observed in Galactic single O-type stars. This is especially the case of the bright 
binary ELST50, which presents the highest \loglxlbol\ 
ratio: even taking into account the presence of two stars, the X-ray 
source is still about 5 times brighter than expected. The ratio appears 
closer to the Galactic value only for PGMW3204/9, considering the sum of the 
individual luminosities of this stellar group's components -- detailed 
stellar content is well known in this area, but maybe not elsewhere in LH10. 
It is also interesting to note that in our data, PGMW3204/9 
appears as bright as PGMW3070, whereas it was half as bright 
during the {\em XMM} observations (compare our Fig.\,\ref{lh10hst} with Fig. 11 
of \citealt{naz04}): this may indicate some variability in its high-energy 
emission, which is not expected for single, `normal' massive stars and is 
therefore generally associated with an orbital or rotational 
modulation. Further monitoring of these objects is needed to ascertain
their nature.

Some diffuse emission is also present throughout the field, especially 
to the southwest of the cluster, near PGMW 3070 and 3120. In this region, 
a wind-blown bubble was detected by \citet{naz01}. However, the same authors found another 
expanding bubble east of PGMW 3204/9 and ELST 50, with similar 
expansion velocity but it does not appear to be associated with diffuse 
emission as bright as the former. Higher velocity structures were also 
detected by \citet{naz01}, but appear to the south and west of the X-ray bright region, not coincident with it. 
It might be noted that the diffuse emission, which appears centrally 
peaked rather than bright-rimmed, corresponds to the region of highest 
stellar density - the eastern part of LH10 being much less crowded. 
This could suggest that at least part of the X-ray emission actually 
comes from unresolved stars, as in the case of NGC602 in the SMC \citep{osk13}. This hypothesis seems confirmed when the 
spectrum of that diffuse emission is analyzed (Table\,\ref{specdif}): 
two thermal components are present, one with a low temperature typical 
of soft diffuse emission and one with higher temperature, most probably 
stellar in origin (both components have similar luminosities in the 0.3--2.0\,keV band).

\subsubsection{Sources in N11A}
The compact HII region N11A, lying to the east of LH10, probably harbors 
the youngest optically-visible stars in the field \citep{hey01}. It is 
associated with the `star' PGMW3264 or ELST28, which is actually the 
earliest component of a compact stellar group composed of 7 objects. The 
associated X-ray emission is rather strong and a comparison of its shape 
with that of its neighbours suggests some extension. Since the earliest 
object was proposed to be a highly obscured mid O-type star \citep{hey01,eva06}, so that strong X-ray emission from that star 
is not expected. The X-ray emission may rather be associated with wind-wind 
interactions between the cluster members or a combination of several 
unresolved sources. Unfortunately, the spectrum of this source is difficult to 
extract since it lies on the edge of some datasets. The crude X-ray 
spectrum only points to a high temperature, 
but the high noise prevents us to draw definitive conclusions.  
These results should therefore be taken with caution (especially since 
a low absorbing column is also favored by the noisy data, clearly 
at odds with the heavy extinction expected within a dense cloud).
N11A deserves further investigation especially to assess its stellar content in detail.

\subsection{Diffuse Emissions in N11}

As can be seen in the color image of the \chandra\ FOV (Fig.\,\ref{fig:rgb}), soft diffuse 
emission pervades the N11 region. On the basis of the surface brightness, 
we defined 11 regions to be further analyzed and three regions displaying lower 
emission were chosen as backgrounds (see Fig.\,\ref{difpos} and 
Table\,\ref{regiondif} for their positions). 

For their study, we need to estimate the non-X-ray background contribution which is not vignetted by the telescope. This estimate uses the ACIS stowed 
background database, which has been processed with charge transfer
inefficiency (CTI) and gain corrections\footnote{following the procedure 
described in http://cxc.harvard.edu/contrib/maxim/acisbg/}.
The background level in each chip of individual observations was further 
normalized according to the ratio of its count rate to that of the 
stowed data in the 10--12\,keV band, where events are almost completely 
due to the non-X-ray background. 

Spectra of the chosen regions were extracted, the non-X-ray 
background being subtracted from each one. Regions of twice the 70\% EER around each point source were removed. Note that for visual presentation of smoothed diffuse X-ray intensity maps, we replace the source-removed region with values interpolated from data in surrounding bins (see e.g. Fig.\,\ref{difpos}). 
The spectra have been analyzed in the 0.3--2\,keV range; beyond which there is little signal.

To take into account the variations of the sensitivity across the 
FOV, 
the spectral fitting was done in two steps. First, each background 
(or combination of background) was individually fitted although the background spectral shape remains rather similar, with the most varying parameter being the overall intensity level.  Second, the sources were fitted by 
models of the form $vphabs*\sum apec+model_{\rm B}$, where the former 
component represents the true spectrum of the diffuse source (metal 
abundances of the absorption and thermal emission models were 
accordingly fixed to 0.3) and the latter represents the best-fit background model
determined before (these were fixed to the best-fit background model with the normalisation
factors scaled by the effective surface ratio - keyword BACSCAL - of the source and background regions). 
The results of these 
fits are presented in Table\,\ref{specdif}. The spectral analysis of 
two additional regions of LH9 that display a higher surface brightness 
is also presented in this table.

The results from spectral fitting allow us to compare the selected regions. First, we examine temperature variations. The spectra of all regions display a thermal component at low temperature 
($kT\sim$0.2\,keV). For LH9 and LH10, an additional, hotter plasma component ($\gtrsim 1$\,keV)
was needed to obtain a good fit, suggesting it to be associated with unresolved stellar 
objects (PMS, O stars). This contamination is not negligible since it 
represents about one-third of the total flux associated with the diffuse 
sources. Its stellar origin is supported by the analysis of 
the region (called `middle') situated between LH9--LH10 and the northern 
limit of the FOV. The spectrum of the latter region does not require the presence of 
a high temperature component: indeed, only a few stars are scattered 
across this region and the stellar contribution is thus expected to be 
negligible here. Note that a fit of the LH9 spectrum where the abundance of 
the low-temperature component is allowed to vary does not improve the 
$\chi^2$.

Regarding absorptions, a value of $N_{\rm H}=7\times10^{21}$\,H-atom\,cm$^{-2}$ is generally found for the 1T fits, and a lower value for the 2T fits. The larger absorption value
might in fact be an artifact: if we use only one component to fit 
the spectra of LH9 or LH10, the best-fit also yields a higher absorption 
for these objects - this may also be linked to the temperature-absorption trade-off mentioned before. Unfortunately, the spectra of the other diffuse sources 
are too noisy for making a meaningful 2$T$ fit. Keeping this caveat in mind, 
one can however see the remarkable homogeneity of the derived spectral 
parameters, with the exception of harder components being needed when stellar clusters are located (LH9, LH10). 

\section{Summary}
This paper reports the first results of a very deep \chandra\ 
observation of the giant HII region N11 in the LMC. Soft diffusion emission is seen throughout the field, but its spectra reveal some point source contamination. Thanks to the long 
exposure ($\sim$300\,ks) and the high spatial resolution, 165 X-ray point-sources 
were detected in the field, with three showing significant temporal 
variability. Our \chandra\ observation thus increases by more than a factor of five the 
number of point-sources known in N11. Keeping in mind that the sensitivity 
varies across the field, it must be noted that the faintest detected sources 
have count rates of about 0.04 cts\,ks$^{-1}$, which correspond to 
luminosities of about 10$^{32}$\,erg\,s$^{-1}$ in the 0.5--8.0\,keV energy band.  
Diffuse emission is also detected throughout the field, but the harder X-ray emission from some regions 
indicates contamination from unresolved stars.

Most of the X-ray sources are background objects seen through the LMC, but there are also 11 Galactic stars. However, 14 OB 
stars are clearly detected in X-rays, three of them corresponding to compact 
clusters (HD32228, PGMW3204/9, and N11A). The known binaries are not preferentially detected in N11, though 
this conclusion might be biased by the incomplete knowledge of the stellar 
multiplicity. Indeed, these stars could correspond to 
interacting winds systems or magnetic objects, as may be suggested by their
rather high luminosities and plasma temperatures. In this context, it should be noted that  changes are detected in massive stars of LH10, compared to older {\em XMM} data. Follow-up observations are therefore needed to ascertain the nature of these sources. The stacked emission of the undetected O-stars yields a \loglxlbol\ of $-7.3\pm0.3$. This suggests that the intrinsic X-ray emission of massive stars could be similar in the Galaxy and the low-metallicity environment of N11. This is unexpected, as X-ray emission from massive stars is known to arise in their line-driven stellar winds whose properties are known to vary with metallicity. Further observations are however needed to confirm this result.

These {\em Chandra} data will be used for several follow-up studies, notably on the SNR N11L (Sun et al., in preparation), the diffuse emission, and a global multiwavelength study of N11.

\acknowledgments

YN acknowledges support from the Fonds National de la Recherche 
Scientifique (FRS-FNRS, Belgium), the University of Li\`ege (incl. ARC), the 
University of Illinois at Urbana-Champaign, and the PRODEX XMM and 
Integral contracts. YHC and RG acknowledge grant SAO G07-8091A, QDW grant SAO G07-8091B and LMO support from BMWI/DLR grant FKZ 50 OR 1302.

{\it Facilities:} \facility{CXO (ACIS)}.

\clearpage


\begin{figure}
\includegraphics[scale=.70]{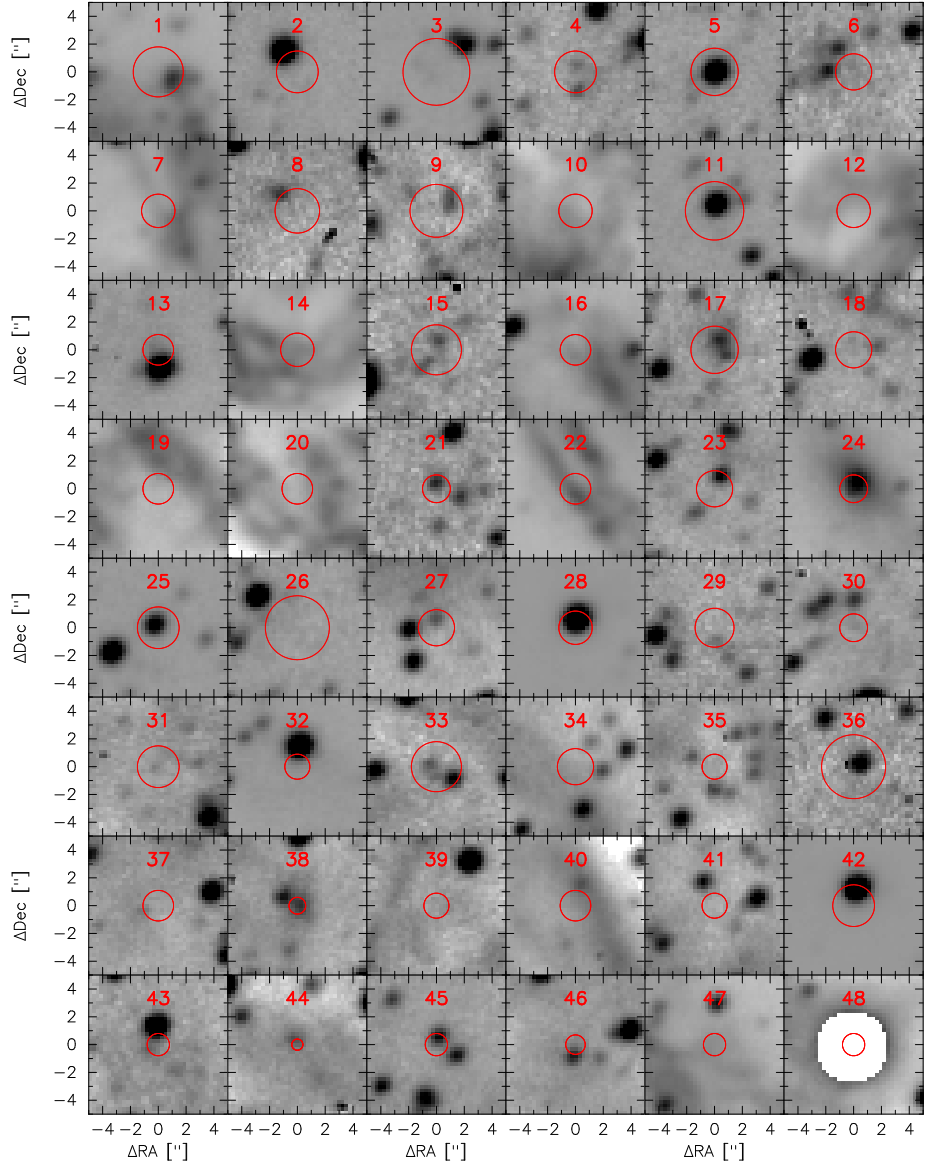}
\caption{{\bf ONLINE MATERIAL} [\ion{S}{2}] images (10\arcsec\ on a side) 
centered on X-ray point sources whose positions are showed by 1$\sigma$-radius 
circular regions. The saturated stellar images appear as white spots in these figures. \label{map}}
\end{figure}
\setcounter{figure}{7}
\begin{figure}
\includegraphics[scale=.70]{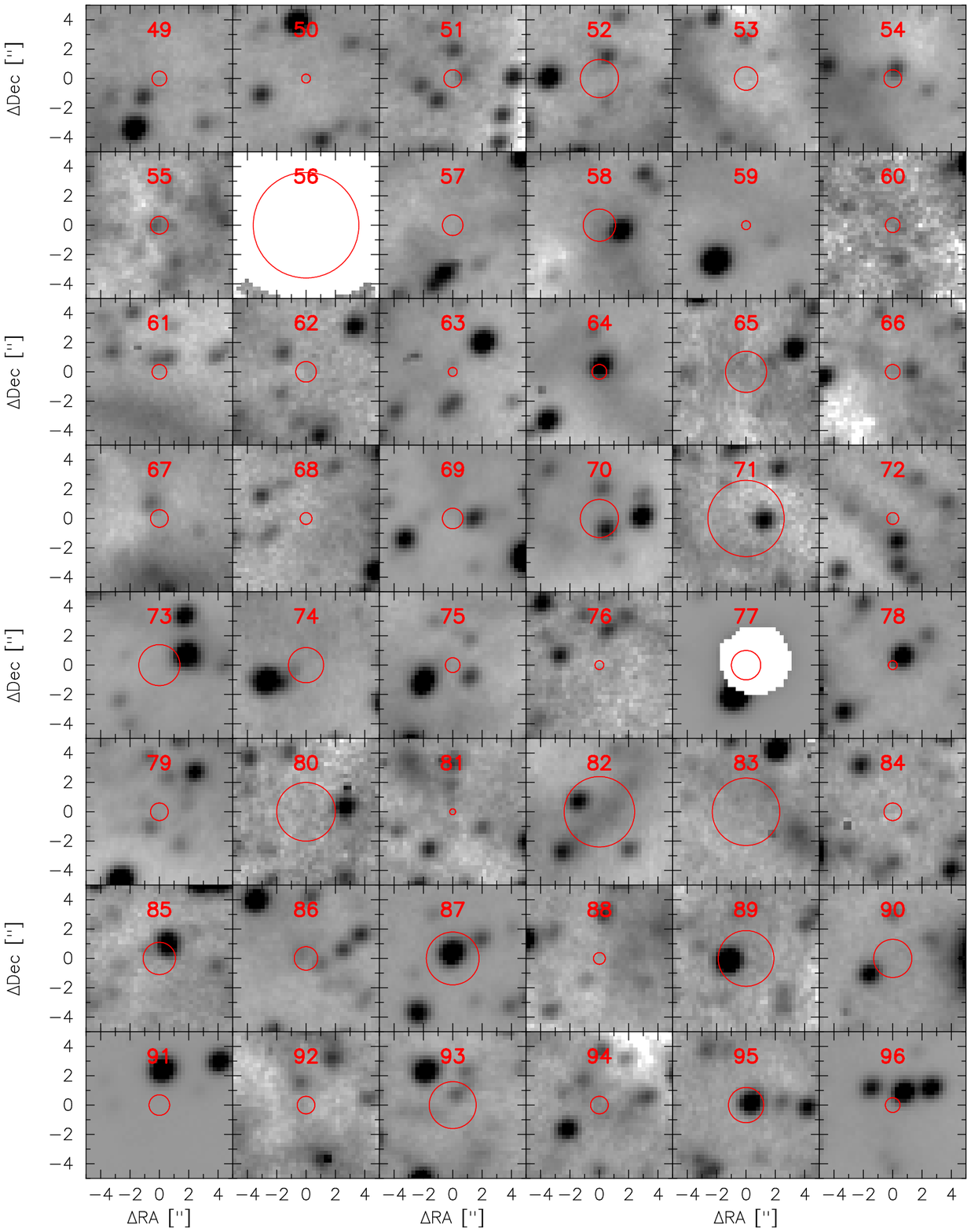}
\caption{Continued}
\end{figure}
\setcounter{figure}{7}
\begin{figure}
\includegraphics[scale=.50]{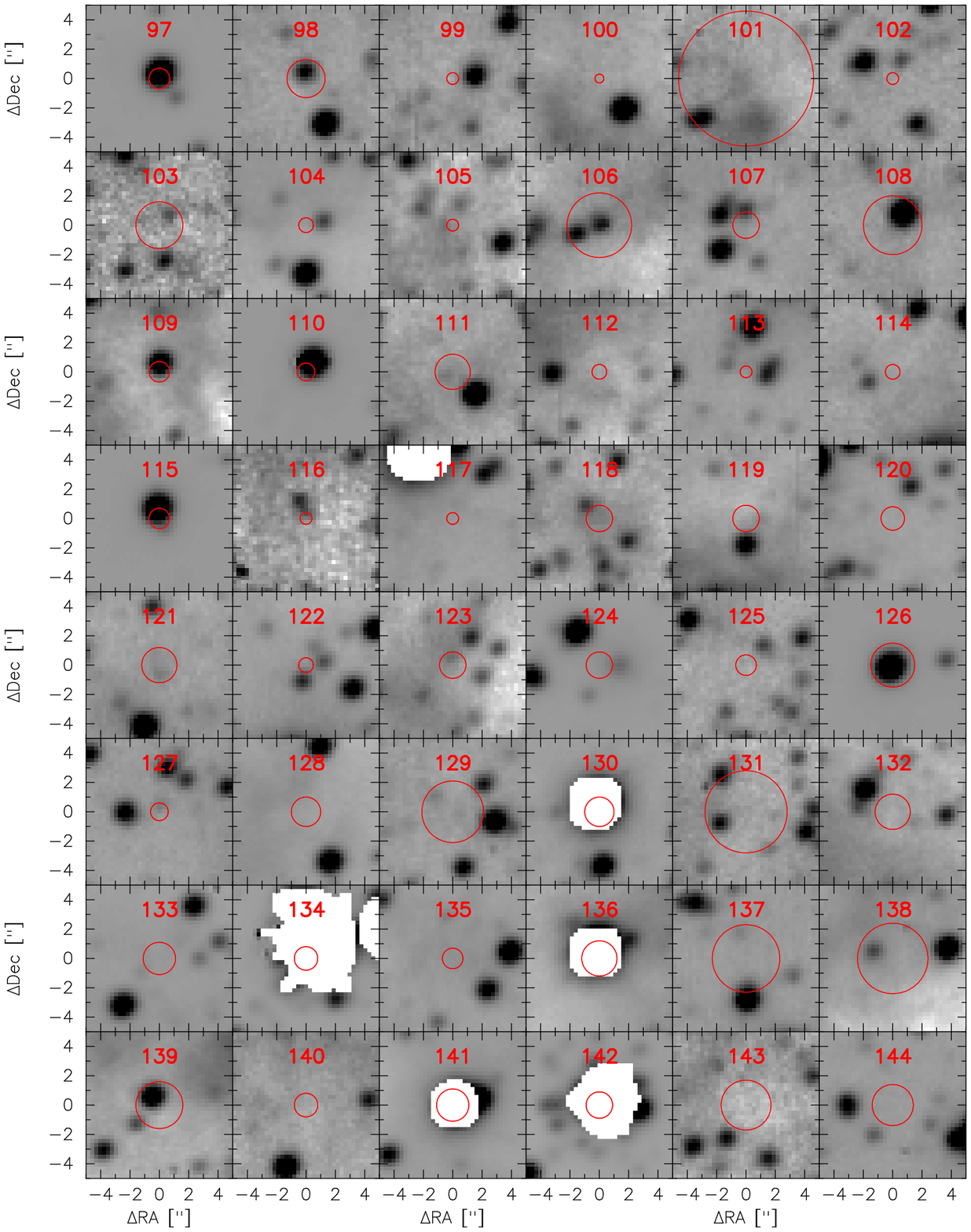}
\includegraphics[scale=.50]{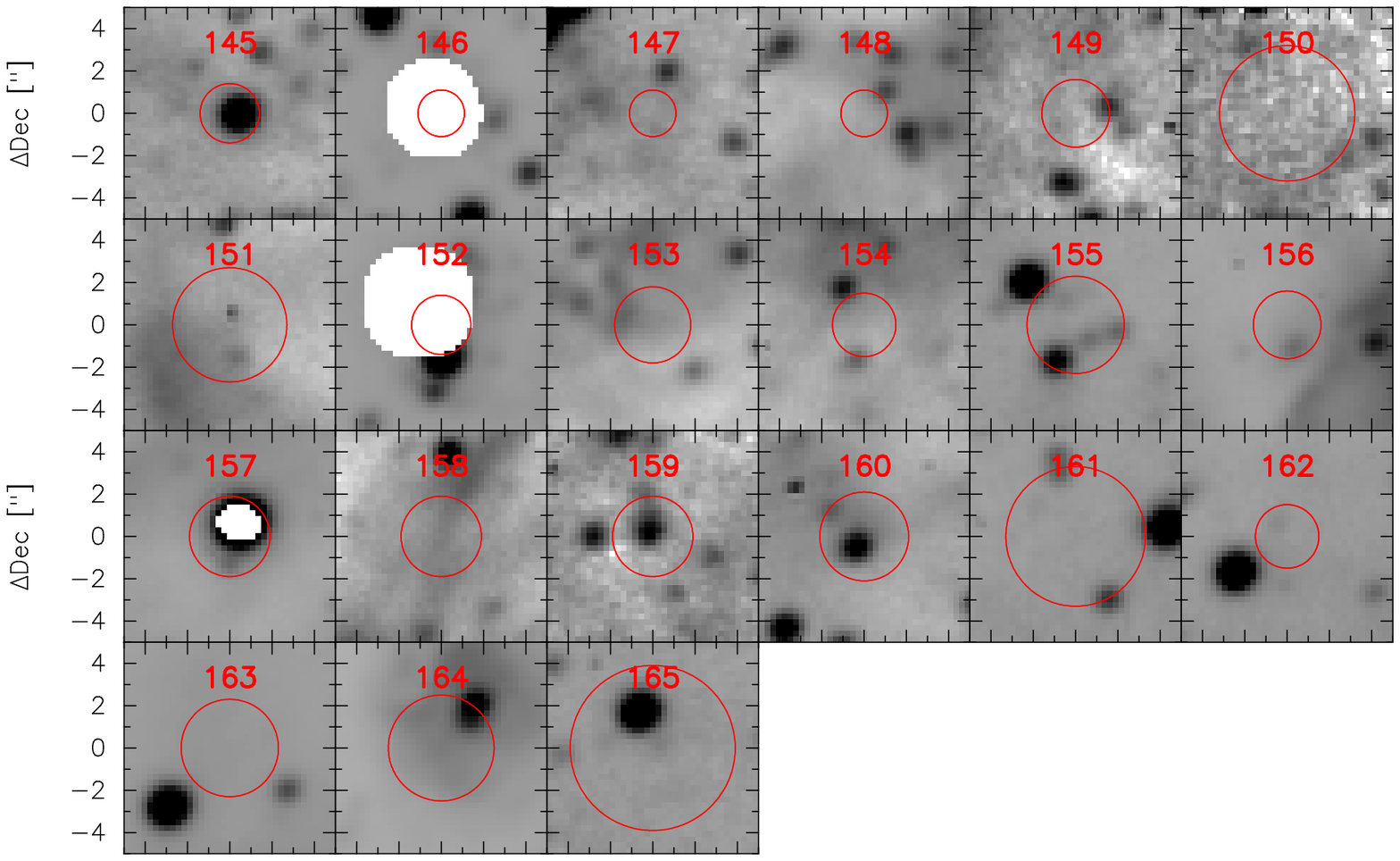}
\caption{Continued}
\end{figure}

\begin{figure}
\includegraphics[scale=.70]{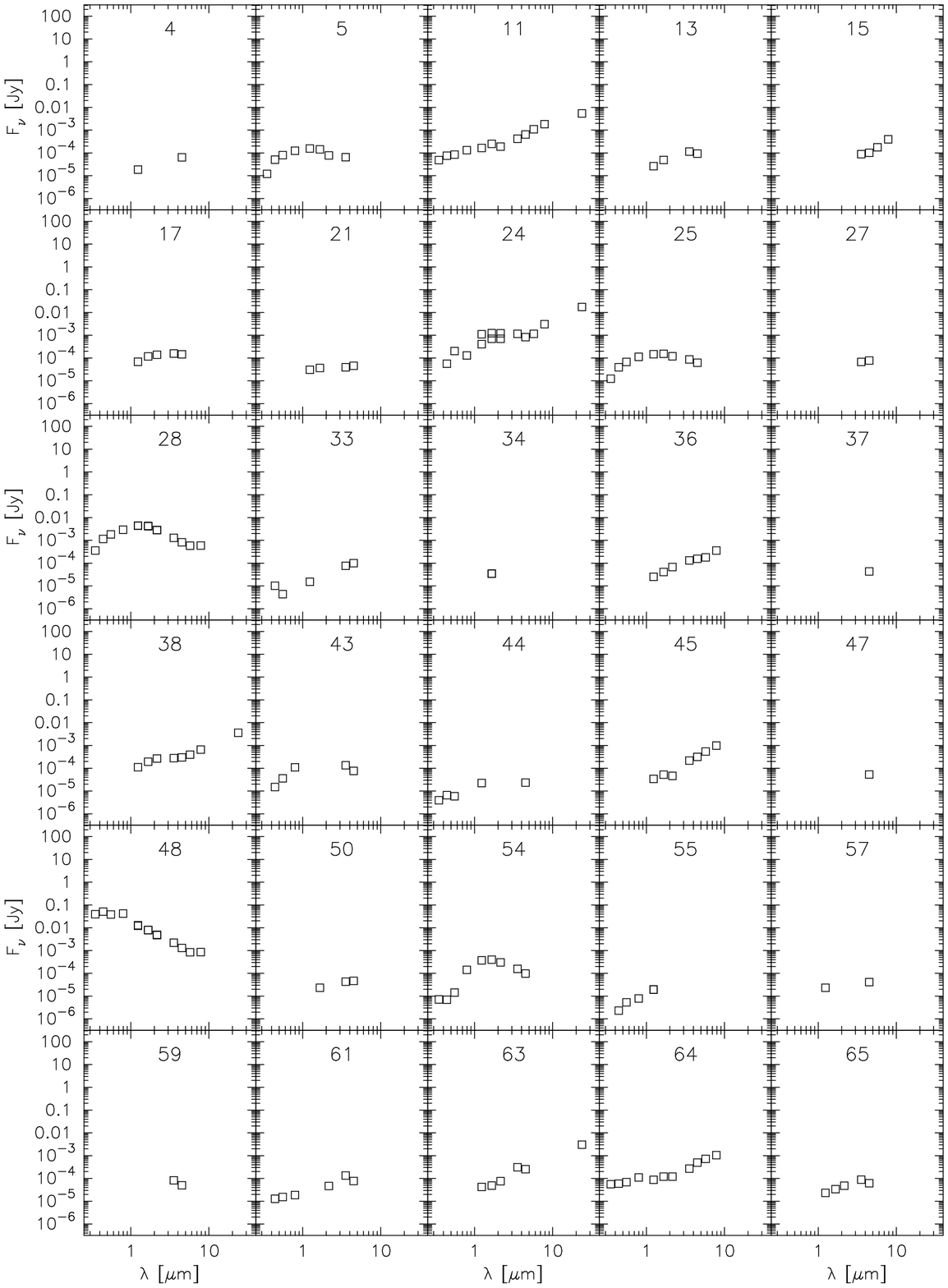}
\caption{{\bf ONLINE MATERIAL} Spectral energy distributions of the counterparts (see text for details). \label{sed}}
\end{figure}
\setcounter{figure}{8}
\begin{figure}
\includegraphics[scale=.70]{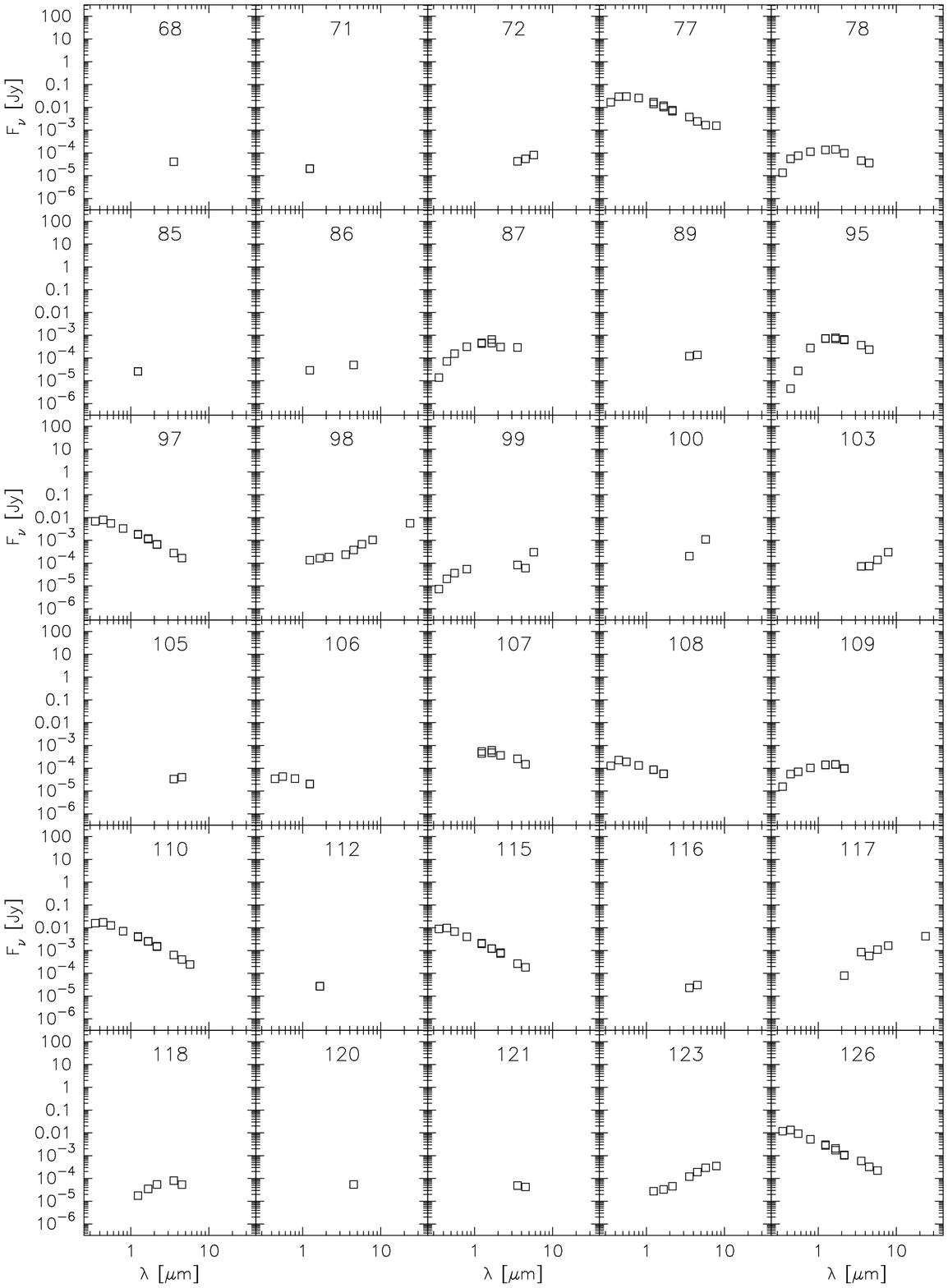}
\caption{Continued}
\end{figure}
\setcounter{figure}{8}
\begin{figure}
\includegraphics[scale=.70]{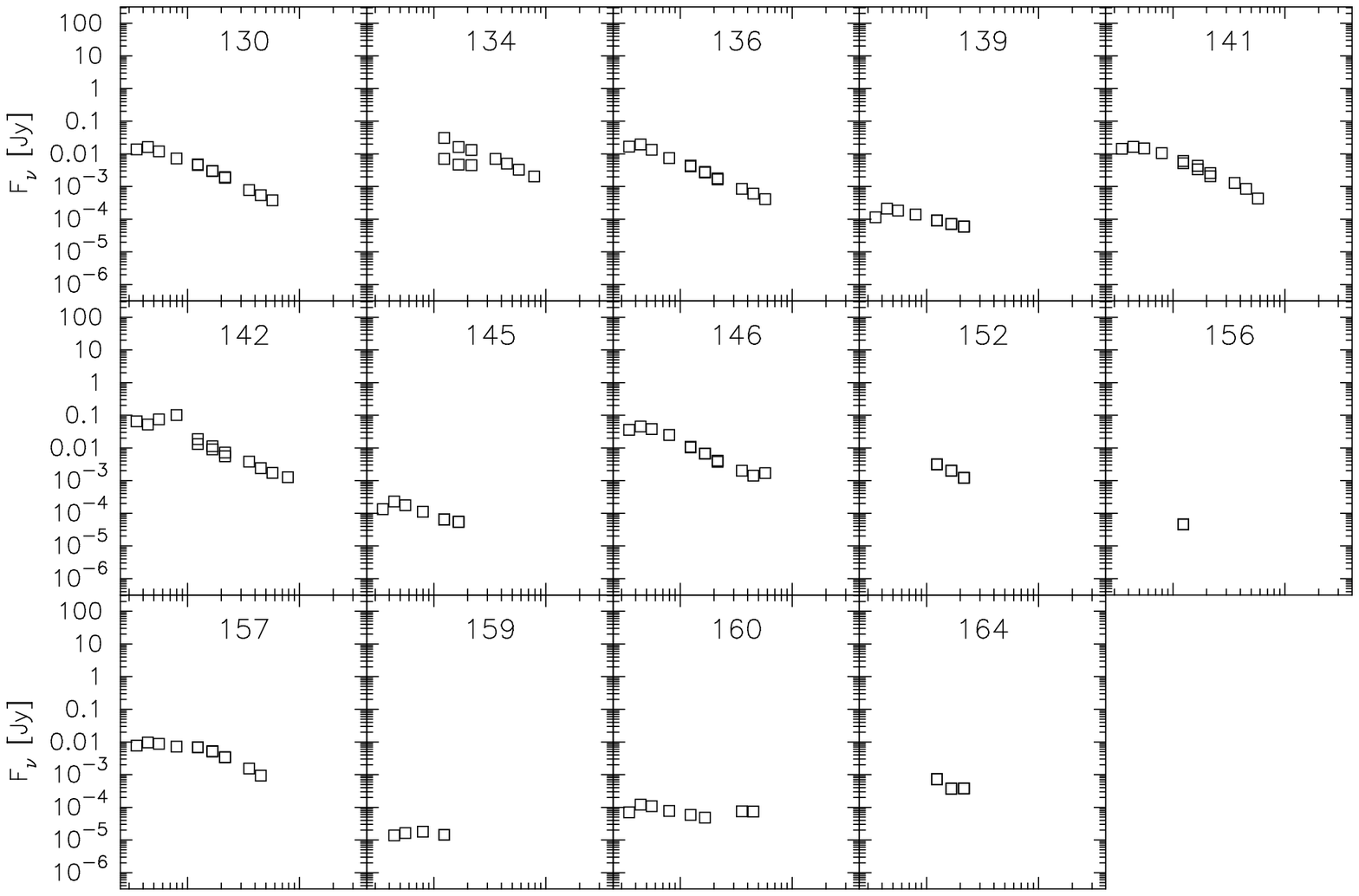}
\caption{Continued}
\end{figure}

\begin{figure}
\includegraphics[scale=.40]{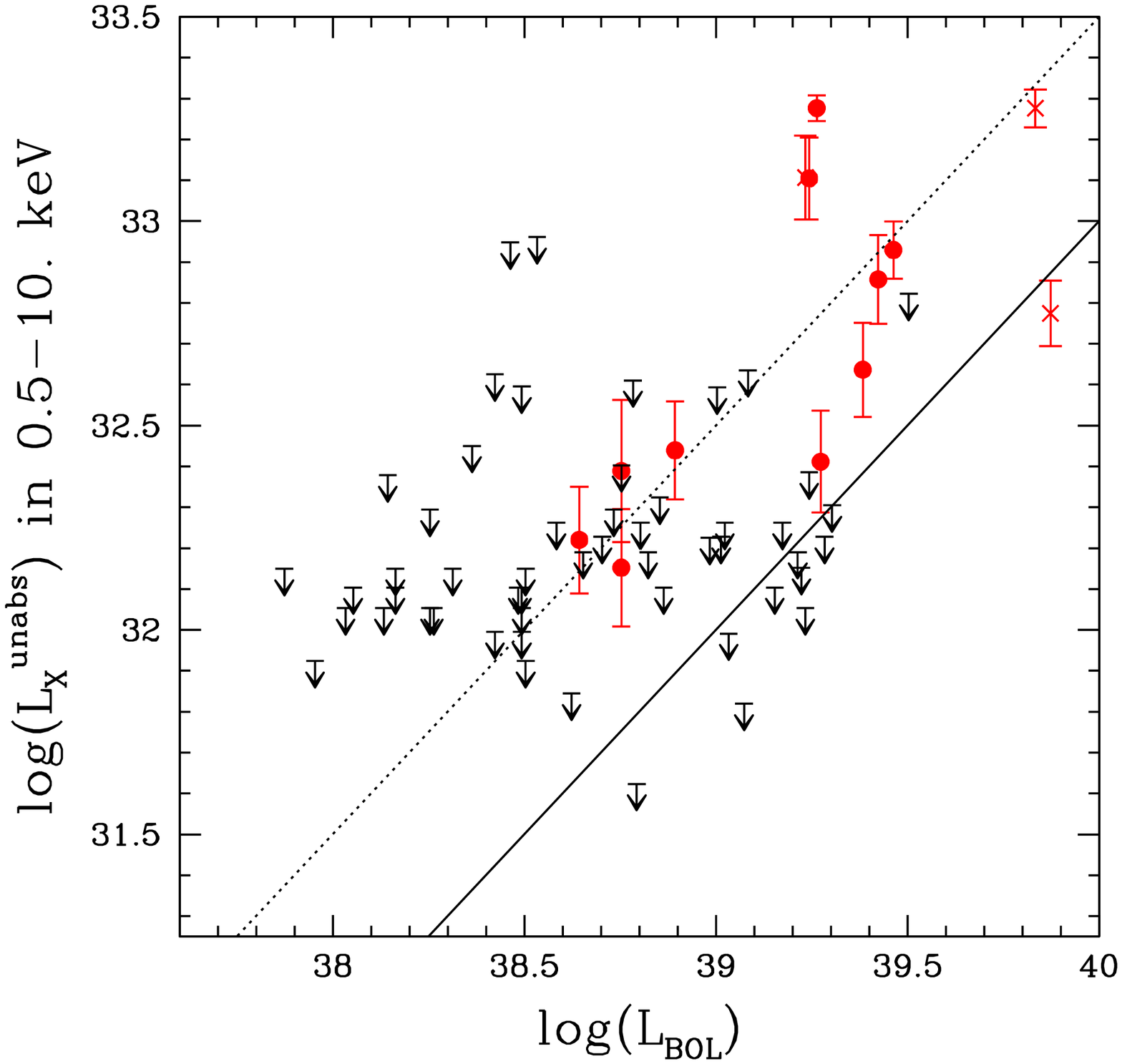}
\caption{$L_{\rm X}-L_{\rm BOL}$ relation for O-type stars in the LMC as derived from the \chandra\ data of N11. The three asterisks and the ten filled circles correspond to the compact groups and other detected O stars, respectively (see Table\,\ref{detO} for details), whereas the arrows give the upper limits on the X-ray luminosity for the undetected objects (see Table\,\ref{allO} for details). The solid line indicates \loglxlbol$=-7.0$, the dotted line is 0.5\,dex higher. \label{lxlbol}}
\end{figure}

\begin{figure}
\includegraphics[angle=-90,scale=.60]{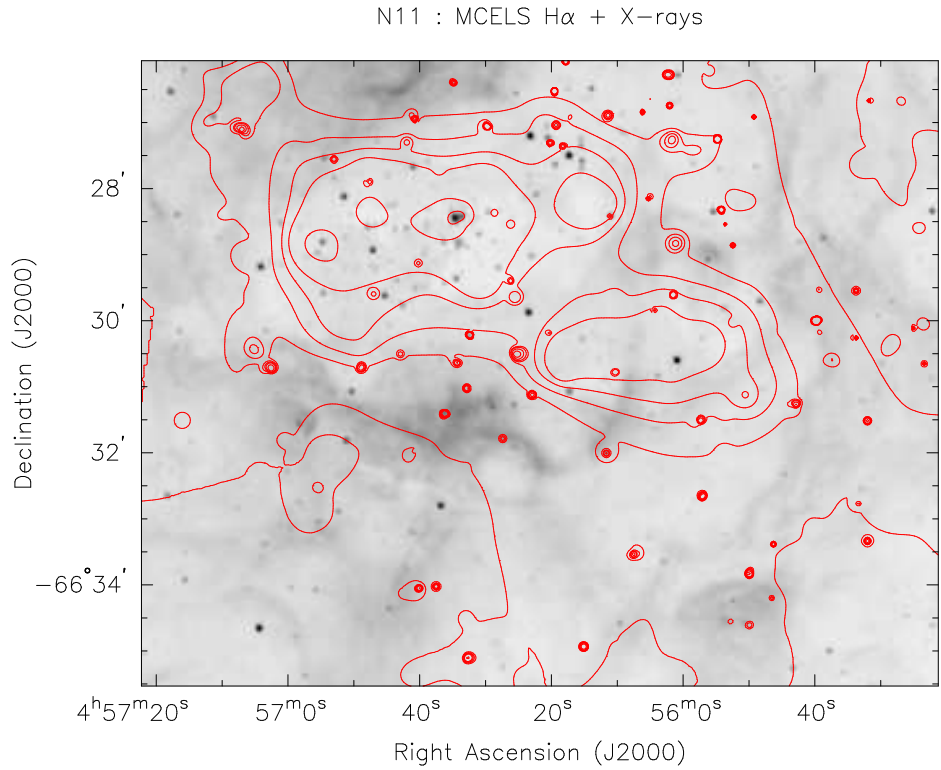}
\caption{H$\alpha$ image from the MCELS with X-ray contours superimposed. The adaptively smoothed X-ray image has been used here, and the contours are set at levels of 0.8,1.5,1.75,2,2.25,3,5 cts\,ks$^{-1}$\,arcmin$^{-2}$. \label{lh9ha}}
\end{figure}

\begin{figure}
\includegraphics[angle=0,scale=.70]{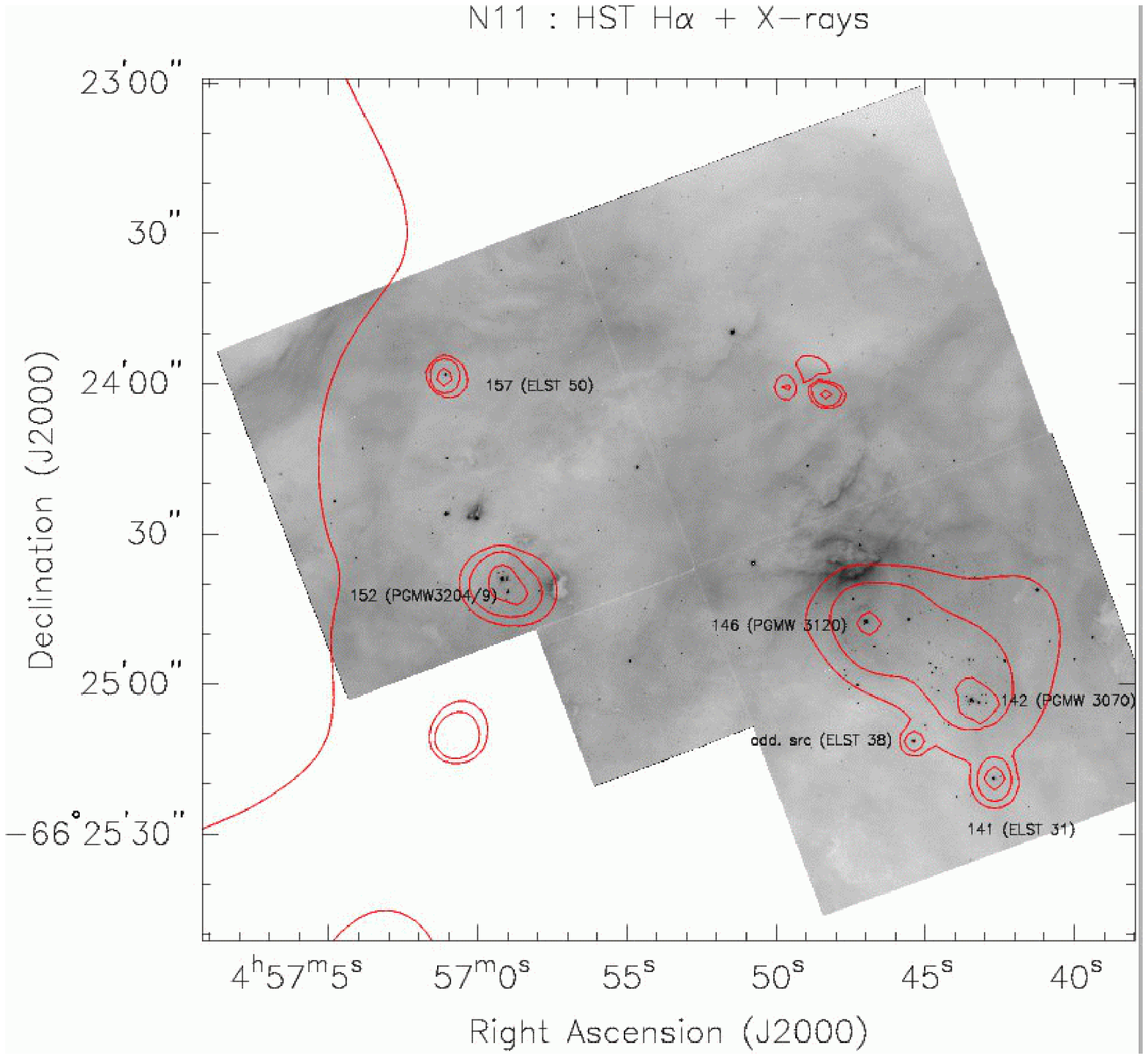}
\caption{Close-up on LH10: {\em HST} H$\alpha$ image \citep[right, from][]{naz01} with X-ray contours superimposed (at levels of 0.8,1.5,2,5 cts\,ks$^{-1}$\,arcmin$^{-2}$). The detected massive stars are labelled by their X-ray source number along with their star name between parentheses. \label{lh10hst}}
\end{figure}

\begin{figure}
\includegraphics[scale=.80]{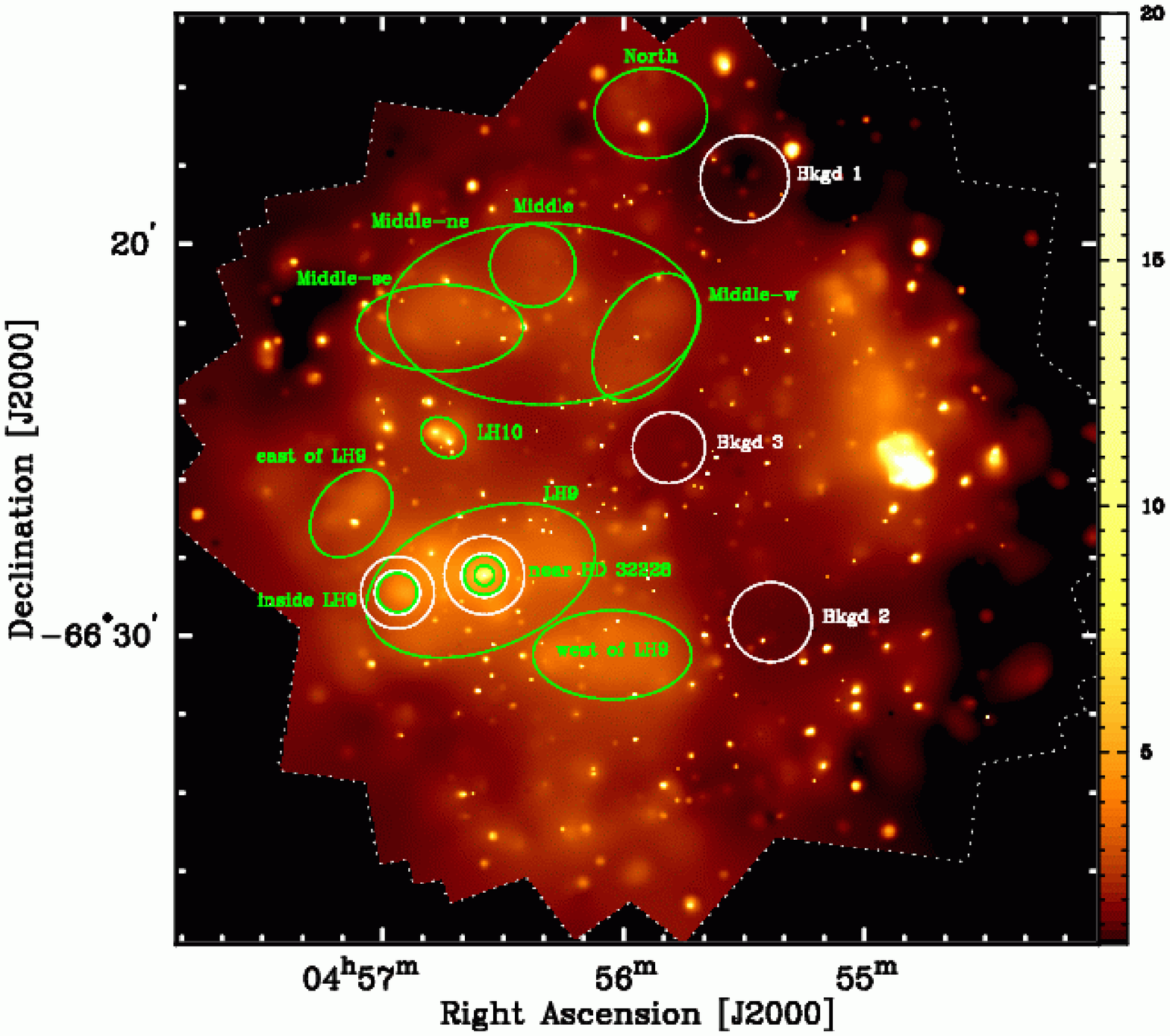}
\caption{{\bf ONLINE MATERIAL} Definition of the extraction regions for diffuse emission (see Table\,\ref{specdif} for their analysis). \label{difpos}}
\end{figure}

\begin{figure}
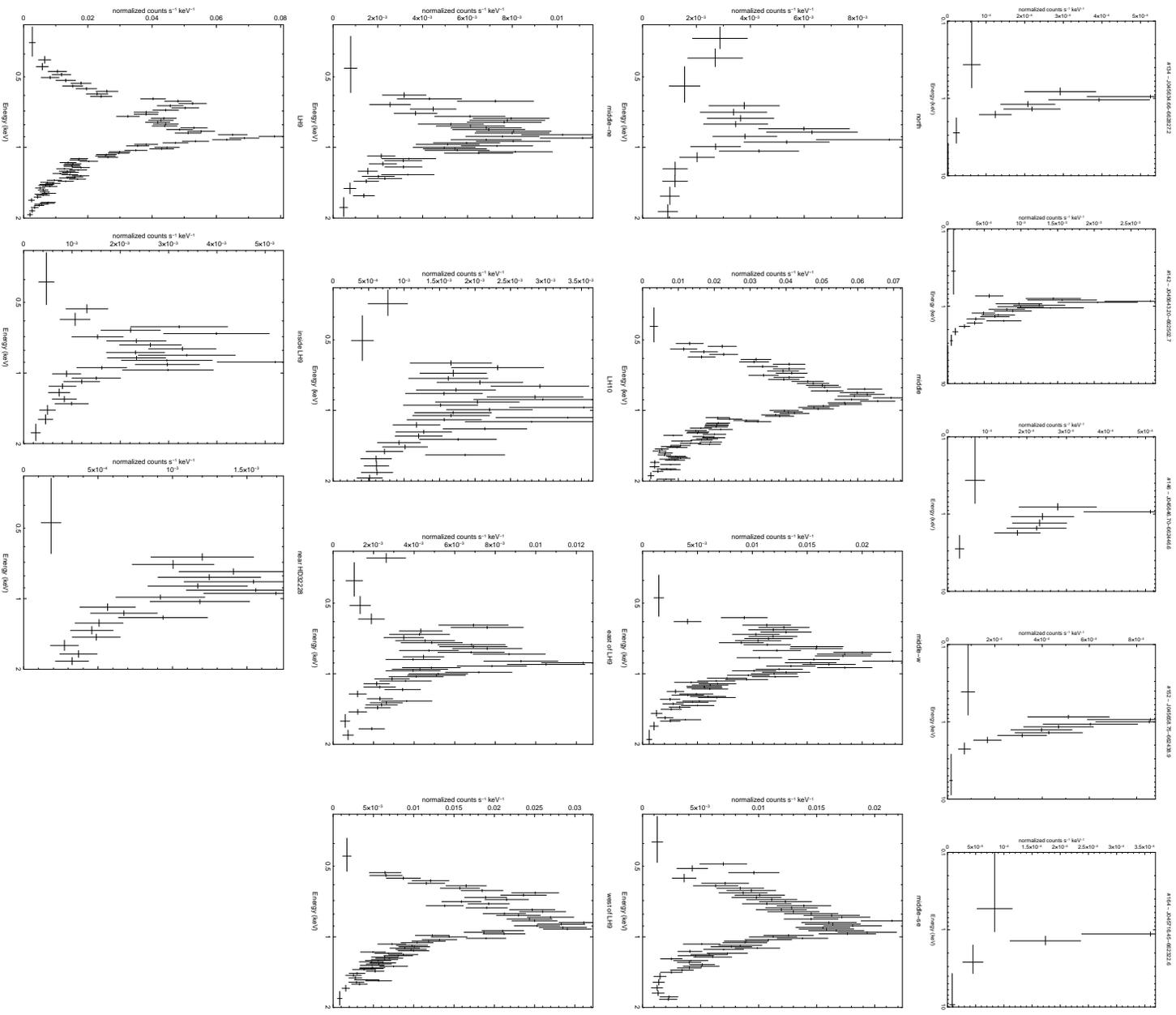

\includegraphics[scale=0.16]{134.ps}
\includegraphics[scale=0.16]{142.ps}
\includegraphics[scale=0.16]{146.ps}
\includegraphics[scale=0.16]{152.ps}
\includegraphics[scale=0.16]{164.ps}
\includegraphics[scale=0.2]{north.ps}
\includegraphics[scale=0.2]{middle.ps}
\includegraphics[scale=0.2]{middle-w.ps}
\includegraphics[scale=0.2]{middle-se.ps}
\includegraphics[scale=0.2]{middle-ne.ps}
\includegraphics[scale=0.2]{LH10.ps}
\includegraphics[scale=0.2]{eastofLH9.ps}
\includegraphics[scale=0.2]{westofLH9.ps}
\includegraphics[scale=0.2]{LH9.ps}
\includegraphics[scale=0.2]{insideLH9.ps}
\includegraphics[scale=0.2]{nearHD32228.ps}
\caption{{\bf ONLINE MATERIAL} Spectra of the point sources (first row, see Table \ref{specO} for fits) and of the diffuse sources (see Table \ref{specdif} for fits). \label{specima}}
\end{figure}

\clearpage

\begin{table}
\caption{{\em Chandra} ACIS-I Observations of N11, by chronological order. \label{obs}}
\begin{tabular}{lcccccc}
\noalign{\smallskip}\noalign{\smallskip}
\tableline\tableline
\noalign{\smallskip}
 &R.A. (J2000) &Dec. (J2000) & Exp\tablenotemark{a} &Roll Ang &OBS Date\\
OBSID  &(h~~m~~s) &($^\circ~~~^{\prime}~~~^{\prime\prime}$)&(s) &($^\circ$) &(yyyy-mm-dd)\\
\tableline
8210     &04 55 50.6 & -66 26 42 &  45486 &   187.7 &2007-05-29 \\
8469     &04 55 50.2 & -66 26 34 &  48895 &   170.1 &2007-06-17 \\
8468     &04 55 50.2 & -66 26 31 &  47359 &   165.1 &2007-06-22 \\
8467     &04 55 50.2 & -66 26 29 &  48383 &   160.1 &2007-06-27 \\
8470     &04 55 50.4 & -66 26 21 &  42492 &   143.8 &2007-07-14 \\
8211     &04 55 50.9 & -66 26 14 &  47359 &   130.2 &2007-07-31 \\
\tableline
\end{tabular}
\tablenotetext{a}{\phantom{0}
The exposure
represents the live time (dead time corrected) of cleaned data.
}
\end{table}

\begin{deluxetable}{lrrrrrrrrrrrr}
 \rotate
  \tabletypesize{\footnotesize}
  \tablecaption{{\bf ONLINE MATERIAL} {\sl Chandra} Source List \label{sourcelist}}
  \tablewidth{0pt}
  \tablehead{
  \colhead{Source} &
  \colhead{CXOU Name} &
  \colhead{$\delta_x$ ($''$)} &
  \colhead{CR (cts\,ks$^{-1}$)} &
  \colhead{HR1} &
  \colhead{HR2} &
 \colhead{$\theta$} &
 \colhead{Cts} &
 \colhead{Bkgd Cts} &
 \colhead{Exp (ks)} &
  \colhead{Flag} \\
  \noalign{\smallskip}
  \colhead{(1)} &
  \colhead{(2)} &
  \colhead{(3)} &
  \colhead{(4)} &
  \colhead{(5)} &
  \colhead{(6)} &
\colhead{(7)} &
  \colhead{(8)} &
  \colhead{(9)} &
  \colhead{(10)} &
  \colhead{(11)} 
  }
  \startdata
   1 &  J045427.48-662518.0 &  1.8 &$     1.05  \pm   0.10$& $-0.17\pm0.13$ & $ 0.53\pm0.13$ &   8.4 &   183 &  42.6 &   191 & T \\
   2 &  J045434.92-662602.1 &  1.5 &$     1.16  \pm   0.10$& $ 0.02\pm0.10$ & $ 0.89\pm0.10$ &   7.6 &   210 &  33.6 &   217 & T \\
   3 &  J045438.74-662207.1 &  2.4 &$     0.21  \pm   0.06$& --& --&   8.5 &    77 &  44.8 &   223 & T \\
   4 &  J045440.30-662408.4 &  1.5 &$     1.72  \pm   0.13$& $-0.39\pm0.09$ & $ 0.52\pm0.09$ &   7.5 &   251 &  35.7 &   178 & T \\
   5 &  J045442.40-662227.4 &  1.7 &$     1.00  \pm   0.09$& $ 0.25\pm0.10$ & $ 0.82\pm0.15$ &   8.0 &   193 &  40.3 &   218 & T \\
   6 &  J045443.47-662518.9 &  1.3 &$     0.66  \pm   0.08$& $ 0.08\pm0.14$ & $ 0.36\pm0.18$ &   6.9 &   121 &  24.1 &   208 & T \\
   7 &  J045444.36-662551.0 &  1.2 &$     1.83  \pm   0.15$& $-0.99\pm0.15$ & $-0.53\pm0.06$ &   6.7 &   213 &  27.7 &   144 & S \\
   8 &  J045444.89-662957.1 &  1.6 &$     0.41  \pm   0.07$& $ 0.34\pm0.17$ & --&   7.3 &    96 &  32.3 &   221 & T \\
\enddata
\tablecomments{The energy bands were defined as follows: 0.5--1.0 (S1), 1.0--2.0 (S2), 2.0--4.0 (H1), and 4.0--8.0\,keV (H2); soft band $S=S1+S2$, hard band $H=H1+H2$, and total band $T=S+H$. 
 Column (1): Running source number. (2): 
{\em Chandra} X-ray Observatory source name, following the
{\em Chandra} naming convention and the IAU Recommendation for Nomenclature
(e.g., http://cdsweb.u-strasbg.fr/iau-spec.html). (3): Position 
uncertainty (1$\sigma$) calculated from the maximum likelihood centroiding and 
an approximate off-axis angle ($r$)
dependent systematic error $0\farcs2+1\farcs4(r/8^\prime)^2$
(an approximation to Fig.~4 of \citet{fei02}), which are added in
quadrature.  Note that they may be overestimated, see \citet[in particular Fig. 9]{get05}.
(4): On-axis source total count rate --- the sum of the 
exposure-corrected count rates in the four narrow
bands. (5-6): The hardness ratios defined as 
${\rm HR_1}=({\rm H-S2})/({\rm H+S2})$, and ${\rm HR_2}=({\rm S2-S1})/{\rm S}$, 
listed only for values with uncertainties less than 0.2.
(7): Off-axis angle ($\theta$; in units of arcminutes).
(8), (9) and (10): Raw counts, background counts, and effective exposure (ks) in the detection aperture.
(11): The label ``T'', ``S'', or ``H'' mark the band in 
which a source is detected with the most accurate position that is adopted in
Column (2). 
}
  \end{deluxetable}

\begin{deluxetable}{lccccccccccccc}
\rotate
  \tabletypesize{\scriptsize}
  \tablecaption{{\bf ONLINE MATERIAL} Counterparts to the X-ray Sources within 1''. \label{counterparts}}
  \tablewidth{0pt}
  \tablehead{
  \colhead{Src} &
  \multicolumn{2}{c}{2MASS} &
  \multicolumn{2}{c}{DCMC} &
  \multicolumn{2}{c}{GSC} &
  \multicolumn{2}{c}{[HKN2006]} &
  \multicolumn{2}{c}{IRSF} &
  \colhead{MCPS} &
  \multicolumn{2}{c}{USNO} \\
  \noalign{\smallskip}
  \colhead{} &
  \colhead{d} &
  \colhead{name} &
  \colhead{d} &
  \colhead{name} &
  \colhead{d} &
  \colhead{name} &
  \colhead{d} &
  \colhead{name} &
  \colhead{d} &
  \colhead{name} &
  \colhead{d} &
  \colhead{d} &
  \colhead{name}
  }
  \startdata
4   & & & & & & & & & 0.67 & 04544030-6624078 & & & \\  
5   & & & & & 0.79 & S1IW031883 & & & 0.10 & 04544240-6622276 & 0.25 & 0.78 & 0236-0039098 \\  
11  & & & & & 0.89 & S1IW033748 & & & 0.59 & 04544661-6620039 & 0.51 & 0.14 & 0236-0039136 \\  
13  & & & & & & & & & 0.48 & 04544726-6627588 & & & \\  
17  & & & & & & & & & 0.57 & 04544895-6631200 & & & \\  
21  & & & & & & & & & 0.76 & 04545253-6626436 & & & \\  
22  & & & & & & & & & & & & 0.86 & 0235-0043997 \\  
23  & & & & & & & & & & & 0.85 & & \\
\enddata
\tablecomments{For each catalog except MCPS, the distance to the X-ray source (d), in arcsec, is first quoted, followed by the name of the counterpart. The type suggested by \citet{hat06} is also given, between parentheses. \\
A * indicates a source associated with a massive star (see Table \ref{detO})\\
For IRSF, there are additional components within 1'' for \# 24, 36, 38, 77, 106, 123, 164 - only the closest source is mentioned above.}
\end{deluxetable}

\begin{deluxetable}{lcccccll}
\rotate
  \tabletypesize{\scriptsize}
  \tablecaption{{\bf ONLINE MATERIAL} Counterparts to X-ray sources found within 1'' of our dedicated data (optical [\ion{S}{2}] image and 3.6 and 8.0 $\mu$m {\em Spitzer} data), and suggested nature of the sources. \label{addcount}}
  \tablewidth{0pt}
  \tablehead{
  \colhead{Src} &
  \colhead{in [\ion{S}{2}]?} &
  \multicolumn{4}{c}{Spitzer} &
  \colhead{Comment} &
  \colhead{Suggested Nature}\\
  \noalign{\smallskip}
  \colhead{} &
  \colhead{} &
  \colhead{3.6$\mu$m?} &
  \colhead{8.0$\mu$m?} &
  \colhead{d} &
  \colhead{name} &
  \colhead{} &
  \colhead{} 
  }
  \startdata
1	&y	&   	&        &     &                  &point source + diffuse X-ray 	        &Galaxy at z=0.024\\
4	&y	&y	&y	 &0.56 &045440.29-662407.9&SED rises in IR				&(AGN)\\
5	&y	&y	&   	 &0.24 &045442.38-662227.6&	                                        &Galactic K0V Star at 4.5 kpc\\
7	&   	&   	&   	 &     &                  &					        &N11L SNR\\
10	&   	&   	&   	 &     &                  &					        &N11L SNR\\
11	&y	&y	&y	 &0.39 &045446.61-662004.1& 				                &AGN (see Sect. 3.)\\
12	&   	&   	&   	 &     &                  &					        &N11L SNR\\
13	&   	&y	&y	 &0.27 &045447.28-662759.2&SED peaking at ~3$\mu$m			&(AGN)\\
\enddata
\tablecomments{Detections in only two {\em Spitzer} channels are noted, but a few named sources have only detections in the other channels. Some sources are quoted with a visual detection (`y' in third and/or fourth columns) without a name listed for the {\em Spitzer} counterpart: this happens because the automated detection misses some sources, especially faint ones in high background regions.}
\end{deluxetable}

\begin{table}
  \caption{Number of counterparts from Table \ref{addcount} per category. \label{countsum}}
\begin{tabular}{lcccccll}
\noalign{\smallskip}\noalign{\smallskip}
\tableline\tableline
\noalign{\smallskip}
Nature & Number\\
\tableline
Galactic objects & 11\\
massive stars (or st. groups) in LMC & 13\\
AB stars in LMC & 6\\
AGNs or galaxies& 9\\
(AGN) & 19\\
\tableline
\end{tabular}
\end{table}

\begin{deluxetable}{cccccccccc}
\tabletypesize{\scriptsize}
\tablewidth{0pt}
\tablecaption{Properties of the OB stars detected in the X-ray domain\label{detO}}
\tablehead{
\colhead{Source }           & \colhead{ELST}      &
\colhead{PGMW}           & \colhead{Sp. Type}  & \colhead{Bin.?} 
& \colhead{$N^{\rm ISM}_{\rm H}$}  & \colhead{$\log(L_{\rm BOL})$} &
\colhead{$F^{unabs}_{\rm X}(10^{-15}$} & \colhead{$L_{\rm X}(10^{32}$} & \colhead{$\log(L_{\rm X}/L_{\rm BOL})$}\\
\colhead{}           & \colhead{}      &
\colhead{}           & \colhead{}  & \colhead{}  &
\colhead{($10^{22}$cm$^{-2}$)}  & \colhead{($L_{\odot}$)}&
\colhead{erg cm$^{-2}$s$^{-1}$)} & \colhead{erg s$^{-1}$)} &\colhead{}\\
}
\startdata
48  &  004 & -- & OC9.7Ib   & N & 0.48 & 5.80$\dagger\dagger$& $1.5\pm0.4$   & $4.3\pm1.2$ &$-6.75\pm$0.12 \\
77  &  092 & -- & O7V       & Y & 0.91 & 5.17                & $0.8\pm0.3$   & $2.5\pm1.0$ &$-6.36\pm$0.17 \\
97  &  068 & -- & O7V       & N & 0.10 & 5.06$\dagger\dagger$& $0.55\pm0.17$ & $1.7\pm0.5$ &$-6.42\pm$0.13 \\
110 &  033 &1005& B0IIIn    & N & 0.23 & 5.07$\dagger\dagger$& $0.27\pm0.14$ & $0.8\pm0.4$ &$-6.8\pm$0.2   \\
115 &  065 &1027& O6.5V     & N & 0.07 & 5.17$\dagger\dagger$& $0.47\pm0.16$ & $1.4\pm0.5$ &$-6.60\pm$0.14 \\
126 &  051 & -- & O5Vn      & N & 0.05 & 5.31$\dagger\dagger$& $0.9\pm0.3$   & $2.8\pm0.8$ &$-6.45\pm$0.12 \\
134$\dagger$ & -- & --                                                             
                & WR+O      &   & 0.00 & 6.29                & $2.0\pm0.4$*  & $6.0\pm1.1$ &$-7.10\pm$0.08 \\
141 &  031 &3061& ON2III    & N & 0.65 & 5.84$\dagger\dagger$& $2.4\pm0.6$   & $7.2\pm1.8$ &$-6.57\pm$0.11 \\
\enddata
\end{deluxetable}

\begin{table}
\begin{center}
\caption{Spectral properties of the brightest OB stars.   \label{specO}}
\begin{tabular}{lllllll}
\hline\hline
Source            & $N^{add}_{\rm H}$      & $kT$           & norm  & $\chi^2$ (dof)          & $F_{\rm X}$    & $F_{\rm X}^{unabs}$ \\
 & ($10^{22}$cm$^{-2}$) & (keV) & ($10^{-6}$cm$^{-5}$) & & \multicolumn{2}{c}{($10^{-15}$erg cm$^{-2}$s$^{-1}$)}\\
\hline
\vspace*{-2mm}\\
134& $0._{0.}^{1.3}$ & $1.3_{0.7}^{2.0}$ & $2.1_{1.6}^{6.9}$ & 0.88 (6) & 1.99 & 1.99\\
\vspace*{-2mm}\\
142& $0.7_{0.4}^{0.9}$ & $0.67_{0.58}^{0.76}$ & $15.4_{10.8}^{21.9}$ & 1.14 (19) & 5.39& 6.32\\
\vspace*{-2mm}\\
146& $0.9_{0.3}^{1.7}$ & $0.8_{0.6}^{1.0}$ & $7.5_{3.8}^{15.1}$ & 1.88 (6) & 1.93 & 2.84\\
\vspace*{-2mm}\\
152& $0.3_{0.}^{0.8}$ & $1.0_{0.7}^{1.3}$ & $8.0_{5.5}^{14.4}$ & 0.83 (10) & 3.40 & 6.31\\
\vspace*{-2mm}\\
164& $0._{0.}^{1.5}$ & $1.3_{0.4}^{2.5}$ & $4.6_{3.2}^{63.7}$ & 1.84 (2) & 2.16& 4.27\\
\vspace*{-2mm}\\
\hline
\end{tabular}
\tablecomments{ Fitted models have the form $vphabs(N^{\rm ISM}_{\rm H})\times vphabs(N^{add}_{\rm H}) \times apec$ with the interstellar absorption $N^{\rm ISM}_{\rm H}$ listed in Table \ref{detO} and the abundances of all components set to 0.3 times solar. The lower and upper limits of the 90\%  confidence interval are shown as subscripts and superscripts, respectively; fluxes are given in the 0.5--10\,keV band and unabsorbed fluxes are corrected by the interstellar absorption only ; spectra were grouped to achieve a minimum of 10 counts per bin.}
\end{center}
\end{table}

\begin{deluxetable}{llcccccc}
\tabletypesize{\scriptsize}
\tablewidth{0pt}
\tablecaption{{\bf ONLINE MATERIAL} Properties of the undetected O-type stars known in the field. \label{allO}}
\tablehead{
\colhead{ELST} & \colhead{PGMW} & \colhead{Sp. type} & \colhead{Bin.?} & \colhead{$\log(L_{\rm BOL}/L_{\odot})$} & \colhead{$N_{\rm H}$}& \colhead{$L^{unabs}_X$} & \colhead{$\log(L_{\rm X}/L_{\rm BOL})$}\\
\colhead{} & \colhead{} & \colhead{} & \colhead{} & \colhead{} & \colhead{(10$^{22}$ cm$^{-2}$)}& \colhead{($10^{32}$erg s$^{-1}$)} & \colhead{}
}
\startdata
007  &     & O8 Ib (f)         &N  & 5.72   &  0.58  &   $<$2.02      &$<-$7.00  \\ 
010  &1310 & O9.5 III + B1-2:  &SB2& 5.42   &  0.29  &   $<$3.92      &$<-$6.41  \\ 
011  &     & OC9.5 II          &SB1& 5.49   &  0.43  &   $<$0.66      &$<-$7.25  \\ 
013  &3223 & O8 V              &SB2& 5.66   &  0.48  &   $<$2.43      &$<-$6.86  \\ 
018  &3053 & O6 II(f$^+$)      &N  & 5.64   &  0.43  &   $<$1.42      &$<-$7.07  \\ 
019  &     & O8-9 III-V((f))   &SB2& 5.40   &  0.24  &   $<$1.68      &$<-$6.76  \\ 
020  &     & O5 I(n)fp         &Y  & 5.50   &  0.14  &   $<$4.32      &$<-$6.45  \\ 
022  &     & O6.5 II(f) var    &N  & 5.45   &  0.29  &   $<$0.98      &$<-$7.04  \\
\enddata
\tablenotetext{*}{The bolometric luminosity comes from the fits of \citet{mok07}.}
\end{deluxetable}

\begin{deluxetable}{lcccccc}
\tabletypesize{\scriptsize}
\tablewidth{0pt}
\tablecaption{Definition of regions used for extracting spectra of diffuse X-ray sources. 
\label{regiondif}}
\tablehead{
\colhead{Name}  & \colhead{RA} & \colhead{DEC} & \colhead{Semi-axes} & \colhead{Pos. angle} & \colhead{Counts} & \colhead{Net count rate}\\
\colhead{} & \colhead{(hh:mm:ss)}& \colhead{($^{\circ}$:\arcmin:\arcsec)} & \colhead{(\arcsec$\times$\arcsec)}& \colhead{($^{\circ}$)} & \colhead{} & \colhead{$10^{-3}$\,cts\,s$^{-1}$}
}
\startdata
north         &04:55:53.2&$-$66:16:39.4& 84.1$\times$69.4 &   0 &1165 & $1.5\pm0.3$ \\
middle        &04:56:19.9&$-$66:21:46.2&233.2$\times$138.7&   0 &17570& $18.1\pm0.6$  \\
middle-w      &04:55:54.6&$-$66:22:21.8&106.6$\times$66.1 & 58.9&4158 & $5.1\pm0.3$ \\
middle-se     &04:56:45.7&$-$66:22:07.9&124.0$\times$66.4 &   0 &4799 & $6.1\pm0.3$ \\
middle-ne     &04:56:22.6&$-$66:20:32.4& circle(63.7)     &     &2314 & $2.4\pm0.2$ \\
LH10          &04:56:44.8&$-$66:24:56.2& 36.1$\times$28.2 &322.9&752  & $1.44\pm0.12$ \\
east of LH9   &04:57:07.8&$-$66:26:52.2& 53.1$\times$73.8 &323.7&2066 & $2.2\pm0.2$ \\
west of LH9   &04:56:02.8&$-$66:30:28.9&118.1$\times$67.9 &   0 &5970 & $8.8\pm0.3$ \\
\enddata
\tablecomments{Counts: number of counts in spectra in the 0.3--2.0\,keV energy band (without background correction). Net count rate: count rate in the 0.3--2.0\,keV energy band for the source region, after correction for the background contributions (non X-ray and local ones) or count rate in the 0.3--2.0\,keV energy band for the background region, after correction for the non X-ray background contribution.\\
$*$ The spectrum of the X-ray point source associated with HD32228, extracted in the 5\arcsec\ region surrounding the star, was presented in Table\,\ref{specO} while the region ``near HD32228'' corresponds to an annulus of inner radius 5\arcsec\ and outer radius 15\arcsec, and the spectrum for ``LH9'' exclude a region of 15\arcsec\ around HD32228, i.e. those are 3 distinct regions. }
\end{deluxetable}

\begin{deluxetable}{lcccc}
\tabletypesize{\scriptsize}
\tablewidth{0pt}
\tablecaption{{\bf Fluxes} of the diffuse X-ray sources. 
\label{specdif}}
\tablehead{
\colhead{Name}  & \colhead{Bkgd} & \colhead{$F_{\rm X}$}    &  \colhead{$F_{\rm X}^{unabs}$} & \colhead{Surf. Bgt}    \\
\colhead{} & \colhead{} & \colhead{($10^{-14}$} & \colhead{erg cm$^{-2}$s$^{-1}$)} & \colhead{($10^{-18}$ erg}\\
\colhead{} & \colhead{} &\colhead{} & \colhead{} &\colhead{cm$^{-2}$s$^{-1}$arcsec$^{-2}$)}\\
}
\startdata
north         & 1   & 1.95 & 1.95 & 0.14\\
middle        & 1+3 & 15.4 &295.3 & 3.70\\
middle-w      & 1+3 & 4.15 & 62.4 & 3.55\\
middle-se     & 1+3 & 4.76 & 73.6 & 3.70\\
middle-ne     & 1+3 & 1.89 & 52.7 & 5.16\\
LH10          & 3   & 0.98 & 2.36 & 1.08\\
east of LH9   & 2+3 & 1.77 & 26.0 & 2.70\\ 
west of LH9   & 2+3 & 7.62 & 154. & 7.74\\
LH9           & 2+3 & 24.3 & 24.4 & 0.53\\
inside LH9   & A    & 0.26 & 0.26 & 0.12\\
near HD32228 & B    & 0.23 & 1.35 & 2.71\\
\enddata
\tablecomments{Fluxes are given in the 0.3--2\,keV band and unabsorbed fluxes are corrected by the full absorption ; surface brightnesses correspond to unabsorbed fluxes divided by the actual area of the region ($BACKSCAL$ keyword). For details on fitting, see text.
}
\end{deluxetable}

\end{document}